\documentclass[prc,reprint,superscriptaddress]{revtex4-1} 

\usepackage{graphicx}
\usepackage{amsfonts}
\usepackage{amssymb}
\usepackage{amsmath}
\usepackage{epstopdf}
\usepackage{colortbl} 
\usepackage{multirow} 
\usepackage{lineno} 
\usepackage{setspace} 
\usepackage{diagbox} 
\usepackage[tight]{units} 

\newcommand{\un}[1]{\ensuremath{\ \mathrm{#1}}} 

\usepackage[pdftex]{hyperref} 

\begin{document}

\title{Ultracold neutron production and up-scattering in superfluid helium between 1.1~K and 2.4~K}
\author{K.~K.~H.~Leung}
\affiliation{Institut Laue-Langevin, BP 156, F-38042 Grenoble, France}
\affiliation{North Carolina State University, Raleigh, NC 27695, USA}
\author{S.~Ivanov}
\affiliation{Institut Laue-Langevin, BP 156, F-38042 Grenoble, France}
\author{F.~M.~Piegsa}
\affiliation{Institut Laue-Langevin, BP 156, F-38042 Grenoble, France}
\affiliation{ETH Z\"urich, Institute for Particle Physics, CH-8093 Z\"urich, Switzerland}
\author{M.~Simson}
\affiliation{Institut Laue-Langevin, BP 156, F-38042 Grenoble, France}
\author{O.~Zimmer}
\affiliation{Institut Laue-Langevin, BP 156, F-38042 Grenoble, France}
\date{\today}

\begin{abstract}
Ultracold neutrons (UCNs) were produced in superfluid helium using the PF1B cold neutron beam facility at the Institut Laue-Langevin. A 4-liter beryllium-coated converter volume with a mechanical valve and window-less stainless steel extraction system were used to accumulate and guide UCNs to a detector at room temperature. At a converter temperature of 1.08~K the total storage time constant in the vessel was $(20.3\pm1.2)\un{s}$ and the number of UCNs counted after accumulated was $91,\!700 \pm 300$. From this, we derive a volumetric UCN production rate of $(6.9 \pm 1.7)\,\mathrm{cm^{-3}\,s^{-1}}$, which includes a correction for losses in the converter during UCN extraction caused by the short storage time, but not accounting for UCN transport and detection efficiencies. The up-scattering rate of UCNs due to excitations in the superfluid was studied by scanning the temperature between 1.2$\,$--$\,$2.4~K. Using the temperature-dependent UCN production rate calculated from inelastic neutron scattering data, the only UCN up-scattering process found to occur was from two-phonon scattering. Our analysis for $T<1.95\un{K}$ rules out the contributions from roton-phonon scattering to $< 29\%$ (95\% C.I.) and from 1-phonon absorption to $< 47\%$ (95\% C.I.) of their predicted levels.

\end{abstract}

\maketitle

\section{Introduction \label{sec:introduction}}

Ultracold neutrons (UCNs) are neutrons with kinetic energies less than the neutron optical potential $U_{\rm opt}$ of well-chosen materials; for instance, beryllium has $U_\mathrm{Be} = 252\un{neV}$. They can reflect from material surfaces at all incident angles, allowing them to be stored in a vessel and studied for times up to a few times the neutron lifetime $(880.3 \pm 1.1 )\un{s}$ \cite{Olive2014}. Due to their low velocities ($v \sim 5\un{m\,s^{-1}}$) and the long times they can spend in media or fields chosen by an experimenter, UCNs have become a valuable tool for high-precision studies of the fundamental properties of the neutron and its interactions. These studies have wide-ranging applications in nuclear physics, particle physics, and cosmology \cite{Nico2005,Abele2008a,Ramsey-Musolf2008a, Dubbers2011}. For instance, they are currently used in searches for the permanent electric dipole moment of the neutron \cite{Baker2006,van-der-Grinten2009, Serebrov2009a, Lamoreaux2009, Altarev2010a, Baker2011,Masuda2012a, Altarev2012,Serebrov2014, *Serebrov2014b}, measurements of the neutron lifetime \cite{Paul2009, Wietfeldt2011, Arzumanov2000a, Serebrov2005, Pichlmaier2010, Huffman2000, Ezhov2005, Salvat2014, Leung2014} and $\beta$-decay correlation parameters \cite{Mendenhall2013, Broussard2013}, and quests for dark matter candidates \cite{Ban2007, Serebrov2008a, Zimmer2010a}, axion-like particles \cite{Baessler2007, Serebrov2010, Jenke2014, Afach2015}, and Lorentz invariance violations \cite{Altarev2009}.

These experiments would benefit from an increased density of UCNs, which has motivated the development of ``next-generation'' UCN sources \cite{Korobkina2007, Frei2007, Anghel2009, Serebrov2009b, Saunders2013, Masuda2012,Lauer2013, Zimmer2011, Piegsa2014, Zimmer2015}. They convert cold neutrons (CNs) to the UCN energy range by allowing them to generate excitations in solid deuterium or superfluid $^4$He \cite{Golub1975,Golub1983}. The latter has the advantage that $^4$He has a zero neutron absorption cross-section and, if the converter is kept at sufficiently low temperatures (typically $\lesssim1\un{K}$), thermal up-scattering of UCNs is sufficiently suppressed. This allows the produced UCNs to survive in the converter for times dominated by wall losses of the vessel, typically $\sim100\un{s}$ \cite{Golub1979,Golub1977}. Since UCNs are not in thermal equilibrium with the converter, densities can be greater than those possible from direct thermal moderation, hence the name ``super-thermal'' production.

The need for more intense UCN sources has motivated the present work of developing a super-thermal helium source with a mechanical UCN valve separating the converter vessel from the extraction guides. This allows UCNs to be accumulated and then used to fill the volume of an external experiment when it is required. A unique feature of our source is the ability to extract UCNs from the converter using a vertical extraction system so that windows, which can cause a loss of UCNs, are avoided \cite{Zimmer2007a}.

Locating the UCN converter volume and its associated cryogenics at the end of a CN beam guide, far from the initial source of neutrons, typically from spallation or fission processes, reduces the needed cooling power and activation of the source. Furthermore, the source can be adapted for each experiment in order to optimize UCN delivery; for instance, the length, volume and coating of the UCN transport guides and converter vessel can be altered.

A successful realization of such a superfluid helium source, which is dedicated to an experiment investigating the neutron quantum levels in Earth's gravitational field \cite{Nesvizhevsky2012}, is described in Refs.~\cite{Zimmer2011,Piegsa2014}. The still large room for improvement has motivated the development of a second UCN source at the Institut Laue-Langevin (ILL), Grenoble, France. The first experiments testing and characterizing this source using the PF1B CN beam facility are presented in this paper.

Inelastic scattering of neutrons \cite{Woods1973,Cowley1971} has been used to reveal the fundamental excitations in superfluid helium \cite{Feynman1956,Feynman1954}. These studies can also be performed by measuring the up-scattering of UCNs caused by interactions between a neutron at rest and excitations in superfluid helium at different temperatures $T$. Using UCNs, one is more sensitive to weak scattering processes due to the intrinsic $1/v$ dependence of the cross section, as well as the potential hundreds of meters path length a UCN can spend in the medium under study \cite{Golub1996}. 

\begin{figure}[htbp]
\begin{center}
\includegraphics[width=0.92\columnwidth]{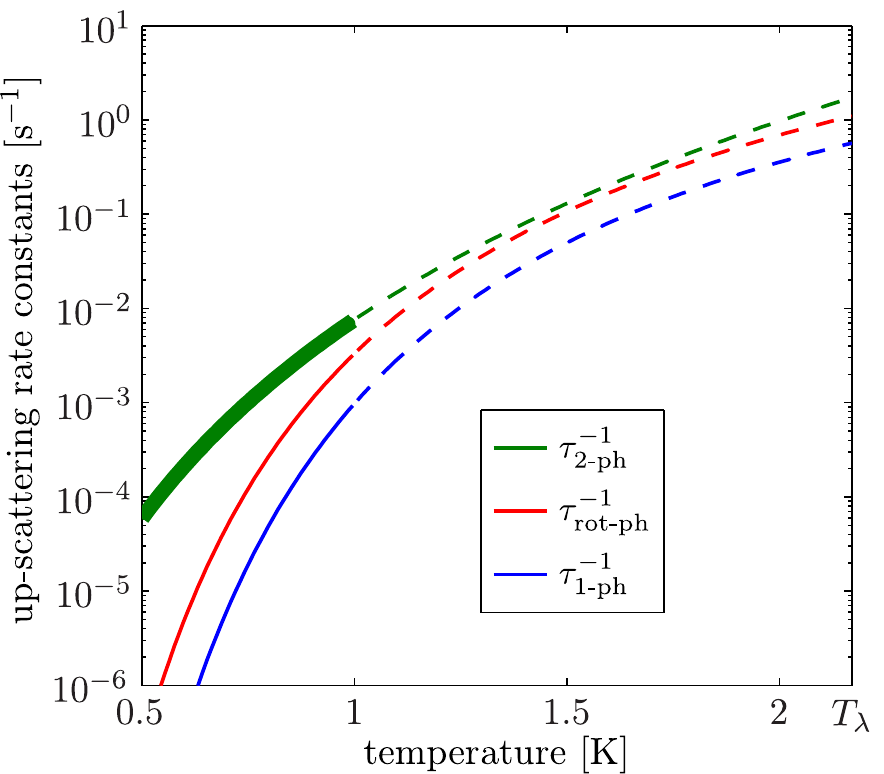}
\caption{(Color online) The sizes of the three main processes contributing to the up-scattering of UCNs by excitations in superfluid helium from Eq.~\ref{eq:tauuptheoretical} derived in Ref.~\cite{Golub1979}. The thickness of the $\tau_\textrm{2-ph}^{-1}$ line covers the range of values for $T< 1\un{K}$ due to the different values of $B$ (see text). The dotted lines are used to indicate that the up-scattering rates are only approximate for $T\gtrsim 1\un{K}$ due to temperature-dependences in the $A$, $B$ and $C$ coefficients.}
\label{fig:strengthOfUpScattering}
\end{center}
\end{figure}

The contributions to the up-scattering rate constant $\tau_{\rm up}^{-1}(T)$ for $T\lesssim 1\un{K}$  have been determined \cite{Golub1979} using the Landau-Khalatnikov Hamiltonian \cite{Khalatnikov1965, Wilks1967} and the three-phonon interaction described in Ref.~\cite{Maris1977} to be
\begin{multline}
\tau_{\rm up}^{-1}(T) = \tau_\textrm{1-ph}^{-1} + \tau_\textrm{2-ph}^{-1}+ \tau_\textrm{rot-ph}^{-1} \\= A\, {\rm e}^{-(\unit[12]{K})/T} + B\,T^7 + C\,T^{3/2}\,{\rm e}^{-(\unit[8.6]{K})/T}\, ,
\label{eq:tauuptheoretical}
\end{multline}
where the first term comes from one-phonon absorption, with $A = 130\un{s^{-1}}$ \footnote{In Ref.~\cite{Yoshiki1992} the original value of $A = 500\un{s^{-1}}$ was used; but the original authors pointed out a new, more appropriate value \cite{Golub1993}.}; the second from 2-phonon scattering (one phonon absorbed and another emitted), with $B = (8.8 \text{ and } 7.6) \times 10^{-3} \un{s^{-1}\,K^{-7}}$ for 0.6\un{K} and 1.0\un{K}, respectively; and the third from roton-phonon scattering (a roton absorbed followed by a phonon emitted), with $C = 18\un{s^{-1}\, K^{-3/2}}$. The $A$, $B$ and $C$ coefficients were calculated for $T\lesssim 1\un{K}$, however, their temperature dependences are weak compared to the overall sizes of the terms. In Fig.~\ref{fig:strengthOfUpScattering}, a plot of these terms up to 2.2\un{K} is shown to illustrate their expected size.

Experiments that have studied the temperature-dependence of $\tau_{\rm up}^{-1}(T)$ have done so up to $1.2\un{K}$ \cite{Golub1983, Kilvington1987, Golub1996}, 1.5\un{K} \cite{Piegsa2014}, and $1.6\un{K}$ \cite{Yoshiki1992}. In all these studies, using $\tau_{\rm up}^{-1}(T) = \tau_\textrm{2-ph}^{-1}$ only was sufficient to describe the data. In the second part of this paper, we study UCN up-scattering to 2.4\un{K}. A detailed data-fitting procedure is used to test the validity of the expected temperature-dependence of $\tau_{\rm up}^{-1}(T)$.

\section{The experimental setup \label{sec:experimentalsetup}}

\begin{figure}[htbp]
\begin{center}
\includegraphics[width=0.95\columnwidth]{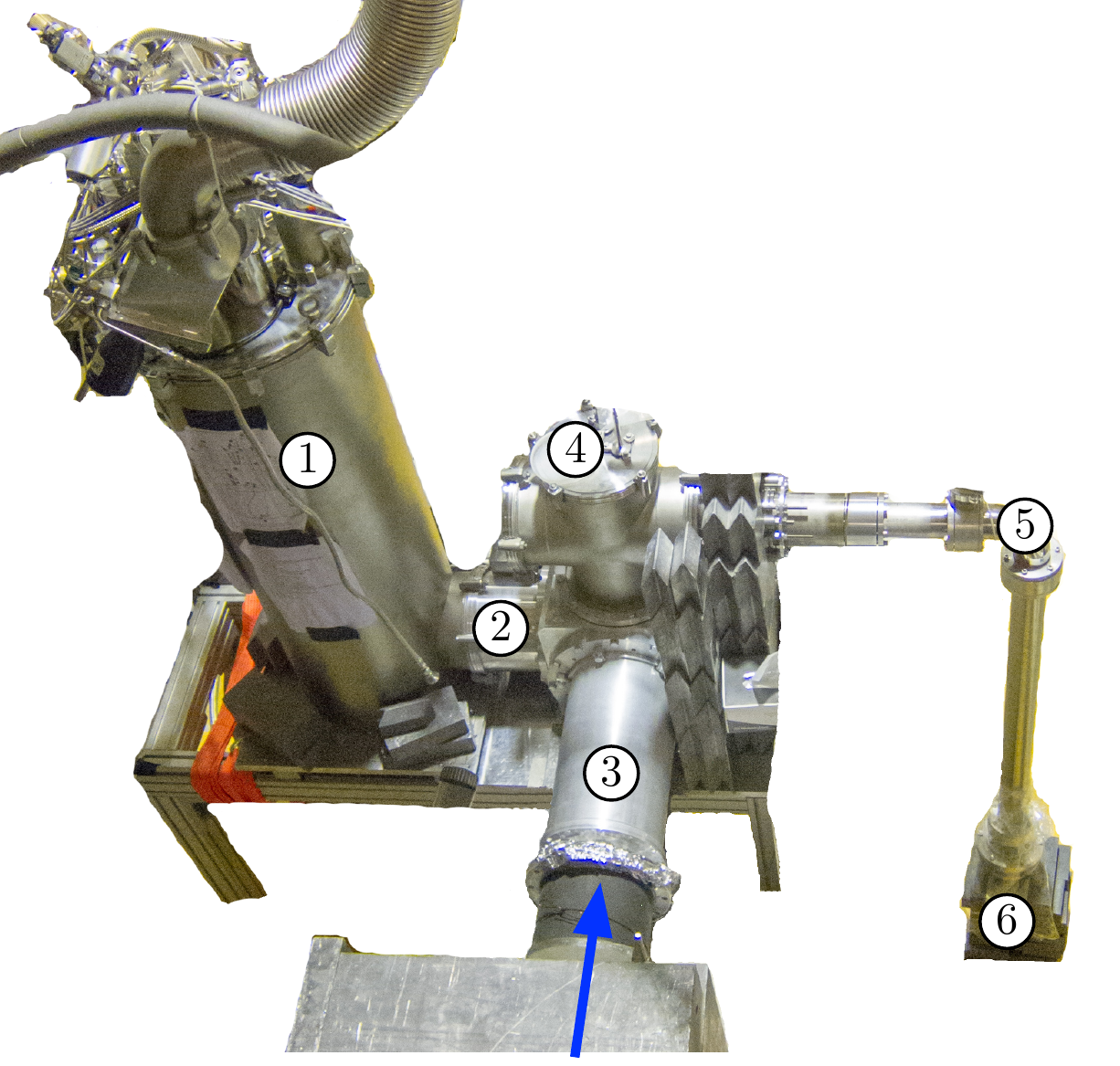}
\caption{(Color online) The experimental installation at the PF1B CN beam line. For clarity, extraneous elements of the original photograph are erased. The added labels indicate:
\textcircled{\protect\raisebox{-0.5pt}{\footnotesize 1}}~main cooling tower;
\textcircled{\protect\raisebox{-0.5pt}{\footnotesize 2}}~horizontal helium connection;
\textcircled{\protect\raisebox{-0.5pt}{\footnotesize 3}}~converter volume;
\textcircled{\protect\raisebox{-0.5pt}{\footnotesize 4}}~mechanical UCN flap valve actuator;
\textcircled{\protect\raisebox{-0.5pt}{\footnotesize 5}}~UCN extraction guides; and
\textcircled{\protect\raisebox{-0.5pt}{\footnotesize 6}}~UCN detector.
The direction of the CN beam is shown as the blue arrow. Additional lead shielding surrounding the apparatus was added later. 
}
\label{fig:experimentOverview}
\end{center}
\end{figure}

The general layout of the experiment installed on the PF1B CN beam line \cite{Abele2006} in 2011, the details of which are described later, is shown in Fig.~\ref{fig:experimentOverview}. A continuous flow $^3$He refrigerator with a measured cooling power of $60\un{mW}$ at $0.6\un{K}$ \footnote{The details of this refrigerator will be described in a separate publication.} is situated in the main cooling tower. This refrigerator is thermally connected, via a copper heat exchanger, to a horizontal stainless steel tube filled with superfluid helium with $50\un{mm}$ diameter and $45\un{cm}$ length. This filled-tube supplies superfluid helium to the converter vessel and also serves for heat transfer. UCNs produced in the helium by the CN beam are extracted vertically from the back of the converter vessel and guided by tubes in a ``$\cap$''-shaped geometry to a UCN detector at room temperature.
 
The superfluid $^4$He used for the converter is purified through a superleak made from Al$_2$O$_3$ powder with a grain size of 50$\,$nm. The fabrication technique is described in Ref.~\cite{Zimmer2010}. Superfluid flow through the superleak is induced via the thermomechanical pumping (or fountain effect) with a heater, similar to that of Ref.~\cite{Yoshiki2005}. The $^3$He to $^4$He isotopic purity from this process is $\lesssim 4\times10^{-10}$ \cite{Atkins1976}.

\begin{figure}[htbp]
\begin{center}
\includegraphics[width=0.95\columnwidth]{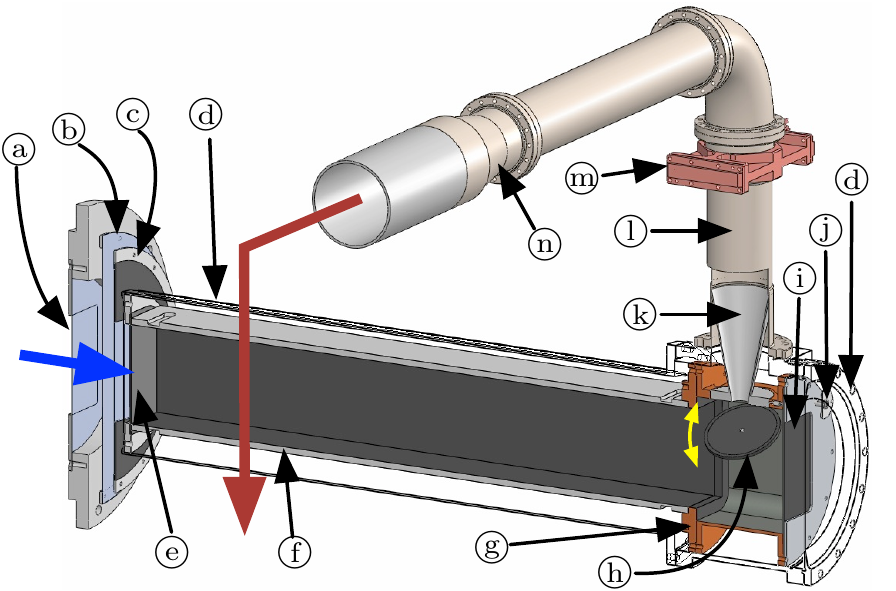}
\caption{(Color online) Cross-section of the superfluid helium converter volume and UCN extraction system.
\textcircled{\footnotesize a}~Isolation vacuum window;
\textcircled{\protect\raisebox{-0.7pt}{\footnotesize b}}~outer radiation shield window;
\textcircled{\footnotesize c}~inner radiation shield window;
\textcircled{\protect\raisebox{-0.6pt}{\footnotesize d}}~superfluid containment vessel (shown transparent);
\textcircled{\footnotesize e}~front beryllium sheet;
\textcircled{\protect\raisebox{-0.7pt}{\footnotesize f}}~beryllium coated aluminum side walls;
\textcircled{\protect\raisebox{+0.7pt}{\footnotesize g}}~beryllium coated copper end section;
\textcircled{\protect\raisebox{-0.7pt}{\footnotesize h}}~beryllium UCN flap valve;
\textcircled{\protect\raisebox{-0.7pt}{\footnotesize i}}~back beryllium sheet;
\textcircled{\protect\raisebox{+0.2pt}{\footnotesize j}}~pulley of UCN flap valve mechanism;
\textcircled{\protect\raisebox{-0.5pt}{\footnotesize k}}~polished stainless steel conical guide;
\textcircled{\protect\raisebox{-0.7pt}{\footnotesize l}}~polished stainless steel UCN guide;
\textcircled{\footnotesize m}~thermal anchoring to 2nd stage radiation shield; and
\textcircled{\footnotesize n}~UCN guide adapter (50\,{\rm mm} to 66\,{\rm mm} ID).
The directions of the incoming CN beam (blue arrow) and the remaining UCN guides to the detector (red arrow) are also shown. (The concentric cylindrical side-walls of the radiation shields and vacuum vessel are not shown.)}
\label{fig:vesselsetup}
\end{center}
\end{figure}

\begin{figure}[htbp]
\begin{center}
\includegraphics[width=0.95\columnwidth]{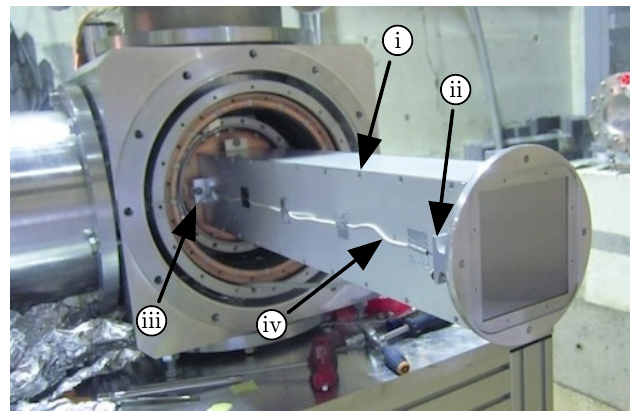}
\caption{(Color online) View of the converter vessel before the superfluid containment vessel is mounted. The added labels and arrows indicate: 
\textcircled{\protect\raisebox{-0.6pt}{\footnotesize i}}~screws for assembling the side walls;
\textcircled{\protect\raisebox{-0.6pt}{\footnotesize ii}}~claw clamps for mounting the front sheet to the side walls;
\textcircled{\protect\raisebox{-0.5pt}{\footnotesize $\mathrm{i\hspace{-0.5pt}i\hspace{-0.5pt}i}$}}~claw clamps for mounting the side walls  to the copper end-section; and
\textcircled{\protect\raisebox{-0.6pt}{\footnotesize $\mathrm{\hspace{+0.7pt}i\hspace{-0.6pt}v}$}}~wiring for thermometer at the front end of converter volume.
}
\label{fig:UCNvolume1}
\end{center}
\end{figure}

The incoming CN beam first passes through front aluminum windows of the vacuum vessel, the two radiation shields, and the superfluid containment vessel before reaching the converter through a 1\un{mm} thick beryllium sheet, which acts as the front window for CNs and the front wall of the converter vessel for UCNs (Fig.~\ref{fig:vesselsetup}). The side walls of the converter vessel are made from four beryllium-coated aluminum plates assembled together with aluminum screws (Fig.~\ref{fig:UCNvolume1}). This assembly is clamped onto a 8$\,$cm long beryllium-coated copper end-section, where the mechanical UCN valve is located. Another 1$\,$mm thick beryllium sheet serves as the back CN window and UCN wall. The internal cross-sections of the converter vessel are $8\times8\,{\rm cm^2}$ in the section made of the aluminum plates and $10\times10\,{\rm cm^2}$ in the copper end section. The volume of the converter vessel with the UCN valve closed is $V_{\rm conv} = 4.0 \un{liters}$. 

The beryllium coatings on the aluminum plates and copper were produced at the Petersburg Nuclear Physics Institute, Russia, by performing two sputtering procedures with intermediate cleaning to reduce defects of coating caused by dust on the surface. Each is estimated to be approximately $400\un{nm}$ thick.

Measurements of the converter temperature are made using 3 calibrated Cernox$^{\rm \tiny TM}$ sensors from Lake Shore Cryotronics placed outside the converter volume but inside the superfluid containment vessel: one is placed at the front of the converter vessel and two at the back on the copper end-section at different heights. During the measurements, the temperatures read at these three sensors were the same to within $< 20\un{mK}$.

\begin{figure}[htbp]
\begin{center}
\includegraphics[width=0.95\columnwidth]{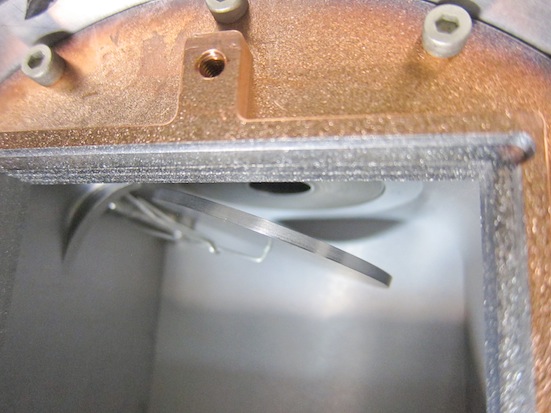}
\caption{(Color online) The mechanical UCN flap valve viewed from the incoming direction of the CN beam. The axle and wires supporting the flap can be seen, as well as the exit for UCNs at the bottom of the stainless steel conical guide section.}
\label{fig:UCNflapvalvephoto}
\end{center}
\end{figure}

The UCN flap valve (Fig.~\ref{fig:UCNflapvalvephoto}) consists of a circular 3$\,$mm thick beryllium disc connected to an axle of a pulley system. A stainless steel cable, driven by a pneumatic piston located outside of the isolation vacuum, is used for closing the valve. A spring, and the weight of the plate itself, provides the force for opening it. The stainless steel axle and wire supporting the plate inside the converter vessel were coated with nickel ($U_{\rm Ni} = 252\,{\rm neV}$) via brush electroplating.

The first section of the vertical UCN extraction guide system is the polished stainless steel ($U_{\rm SS} = 185\un{neV}$) conical guide piece. It reduces the radiative heat transfer into the converter volume from the guides and also aids in aligning the outgoing UCN trajectories vertically so that they can overcome the gravitational potential with fewer reflections off the walls of the guides. The diameter of the entrance to the cone is 23\un{mm}. The rest of the UCN extraction guides, also made of polished stainless steel, has a 50\un{mm} ID until around half-way on the horizontal section after which it increases to 66\un{mm}. The vertical distance from the closed UCN valve to the bottom of the horizontal section is $28\un{cm}$.

The horizontal guides have a total length of $\approx 80\un{cm}$ and the vertical drop to the UCN detector is 1.2\un{m}. The detector is a $^3$He gaseous wire chamber with an aluminum ($U_{\rm Al} = 54\un{neV}$) front window. The dead time of this detector was measured to be $2\un{\mu s}$ \cite{Zimmer2011}.

The conductive heat load down the walls of the guides is reduced by thinning out the straight sections of the guides to a wall thickness of 0.5$\,$mm. In addition, the guides are thermally anchored to the inner and outer radiation shields at some points. A difficulty of using the window-less extraction scheme is the heating due to thermal radiation. Radiation baffles can not be used since they would obstruct UCNs. To reduce this radiative load, the thermal anchoring to the inner heat shield is done both at the copper block and at the 90$^{\circ}$ bend just above it to reduce the temperature of surfaces that have a direct view to the converter. In the current setup, there is no mechanism for suppressing superfluid helium film flow. With a similar extraction guide system and a $^3$He refrigerator with similar cooling power, a converter temperature of $< 0.72\,{\rm K}$ has previously been achieved \cite{Zimmer2011,Zimmer2007a,Zimmer2010}.

The PF1B CN beam line is a 74\un{m} long $m=2$ super-mirror guide coupled to a liquid deuterium cold source \cite{Ageron1989}. The cross-section of the neutron guide in the casemate has a dimension of $6(W) \times 20 (H) \,{\rm cm^2}$. This was cut with an aperture so that two extension guides with cross-sections of $12 \times 5 \, {\rm cm^2}$, forming a combined length of $2\times2\,$m, could be used. The beam was then further cut to $8\times 5\, {\rm cm^2}$ just before the converter vessel. The total volume in contact with the beam (neglecting its divergence within the converter vessel) is $V_{\rm CN} = (2,\!300 \pm 100) \, {\rm cm^3}$. A CN beam monitor detector behind a small circular aperture was placed behind the converter volume outside the cryostat for monitoring the relative CN flux.

An older characterization of the PF1B beam inside the casemate \cite{Abele2006} measured a total capture flux of $1.4\times10^{10}\un{cm^{-2}\;s^{-1}}$, with a spectrum that peaks at 4~\text{\AA} with a differential capture flux of $2\times10^{9}\un{cm^{-2}\;s^{-1}\;\text{\AA}^{-1}}$, and at 8.9~\text{\AA}, where single-phonon UCN production takes place, a differential capture flux of $7\times10^{8}\un{cm^{-2}\;s^{-1}\;\text{\AA}^{-1}}$. In around 2007, after the measurements of Ref.~\cite{Abele2006}, parts of the guide were repaired and the in-pile part upgraded, which increased the total capture flux to $\sim2.2\times10^{10}\un{cm^{-2}\;s^{-1}}$ \cite{Soldner2015}. The capture flux at 8.9~\text{\AA} may have changed more strongly due to colder neutrons making on average more reflections on the guides. The extension guides, despite accepting full divergence of the primary beam, would have also introduced some loss. The estimated flux for UCN production in the converter is only expected to be within $\sim 30\%$ of the above values \cite{Soldner2015}. Since this is used only for comparing the measured UCN production rate in this paper, it is considered to be of sufficient precision.

While the mean-free-path of $8.9\un{\AA}$ neutrons in superfluid helium is $\sim 17\un{m}$, it decreases to $\sim 70\,{\rm cm}$ at $3\un{\AA}$ \cite{Sommers1955}, the center of the broad multi-phonon production peak. A two-component epoxy doped with boron (natural isotopic composition) carbide powder was coated on the cylindrical side-walls of the aluminum radiation shield surrounding the converter volume to absorb scattered neutrons. A lead wall at least $15\,{\rm cm}$ thick in all places was sufficient for biological shielding. 

The background rate at the UCN detector when the CN beam is switched on comes primarily from prompt, scattered CNs. An upper-limit, deduced from having the converter at 2.4~K and UCN valve closed, is $< 12 \un{s^{-1}}$. This rate is higher than typical backgrounds when using a monochromated $8.9\un{\text{\AA}}$ CN beam, common for superfluid helium UCN production. The ambient background rate when the CN beam switched off is $<0.05\un{s^{-1}}$. Further details of the experimental setup and the data described in the subsequent sections can be found in Ref.~\cite{Leung2013}.

\section{The UCN population during measurements}

An accumulation measurement starts with the UCN valve closed and no UCNs in the converter volume. The CN beam is switched on at $t=0$. Assuming the UCNs are in mechanical equilibrium \cite{Golub1991}, if $N(E,T,t)\,{\rm d}E$ is the number of UCNs inside the volume with total energy between $E$ and $E+ {\rm d}E$, the build-up of UCNs will follow
\begin{equation}
\label{eq:differentialaccumulation}
N(E,T,t) = N_{\infty}(E,T)\, \left\{ 1 - \exp[-t/\tau_{\rm tot}(E,T)] \right\} \,
\end{equation}
where 
\begin{equation}
N_\infty(E,T)=  P(E,T) \,V_{\rm CN}  \, \tau_{\rm tot}(E,T)\;.
\label{eq:differentialspectrumequilbrium}
\end{equation}
$P(E,T)$ is the differential UCN production rate per unit time and volume.

The total characteristic storage time constant, $\tau_{\rm tot}(E,T)$, can be written in terms of loss rate constants
\begin{equation}
\tau_{\rm tot}^{-1}(E,T) = \tau_{\rm 0}^{-1}(E) + \tau_{\rm up}^{-1}(T) + \tau_{\rm abs}^{-1} + \tau_{\rm \beta}^{-1}\, ,
\label{eq:storagetimeclosedvolume}
\end{equation}
where $\tau_{\rm 0}^{-1}(E)$ consists of: (1) capture or up-scattering losses through interactions with both the vessel walls (due to the UCN wave-function tunneling into the bulk material of the walls) and impurities frozen on the walls, and (2) UCNs escaping through gaps between walls of the vessel and through the imperfect seal of the UCN valve when it is closed. $\tau_{\rm 0}^{-1}(E)$ is higher for larger $E$ due to an increased wall collision rate, as well as an increase in the loss probability per reflection \cite{Golub1991}. The loss due to up-scattering of UCNs by excitations in superfluid helium, $\tau_{\rm up}^{-1}(T)$, is described in Eq.~\ref{eq:tauuptheoretical}. The loss of UCNs on impurities distributed in the converter, which is primarily due to remaining $^3$He contamination after super-leak filtration, is denoted by $\tau_{\rm abs}^{-1}$ and is $E$- and $T$-independent. Finally, $\tau_{\rm \beta}$ is the neutron $\beta$-decay lifetime.

$P(E,T)$ is dependent on the converter temperature $T$ (see the appendix). The super-thermal down-conversion process is expected \cite{Golub1977} to fill phase-space with a constant density so that $P(E,T) \propto \sqrt{E}$. It should be noted that, since $\tau_{\rm tot}(E,T)$ is shorter for higher UCN energies, the spectrum of accumulated UCNs $N_\infty(E,T)$ is not the same as the one produced via $P(E,T)$. 

After reaching saturation in the accumulation, i.e., $t\gg\tau_{\rm tot}(E,T)$ for all $E$ of UCNs storable in the converter volume, the next step is to switch off the CN beam and then open the UCN valve. Denoting the time when this occurs as $t=t_0$, then the open vessel will empty according to
\begin{equation}
N(E,T,t) = N(E,T,t_0) \,\exp[-(t-t_0)/\tau_{\rm e}(E,T)]\;,
\label{eq:differentialexponentialdecay}
\end{equation}
where
\begin{equation}
\tau_{\rm e}^{-1}(E,T) \approx \tau^{-1}_{\rm tot}(E,T) + \tau_{\rm v}^{-1}(E) \,
\label{eq:emptyingtime}
\end{equation}
is the emptying rate constant. $\tau_{\rm v}^{-1}(E)$ is the rate constant of UCNs escaping through the exit of the opened valve. This is only approximate since it assumes that opening the UCN valve does not change $\tau_0(E)$. Also, since $\tau_{\rm e}(E,T)$ is shorter for higher $E$, the spectrum $N(E,T,t)$ evolves with time.
 
For a continuous measurement, which also starts with no UCNs in the converter volume, the CN beam is switched on with the UCN valve left open. Then the build-up of UCNs (the term accumulation is reserved for when the UCN valve is closed) will follow Eq.~\ref{eq:differentialaccumulation} but with $\tau_{\rm e}(E,T)$ in place of $ \tau_{\rm tot}(E,T)$. After saturation is reached, i.e. $t\gg\tau_{\rm e}(E,T)$, the CN beam is switched off (at $t = t_0$). The population of UCNs in the vessel will empty according to Eq.~\ref{eq:differentialexponentialdecay} as before.

\section{1.08~K Accumulation measurement\label{sec:measurementwithaccum}}

\begin{figure}[htbp]
\begin{center}
\includegraphics[width=\columnwidth]{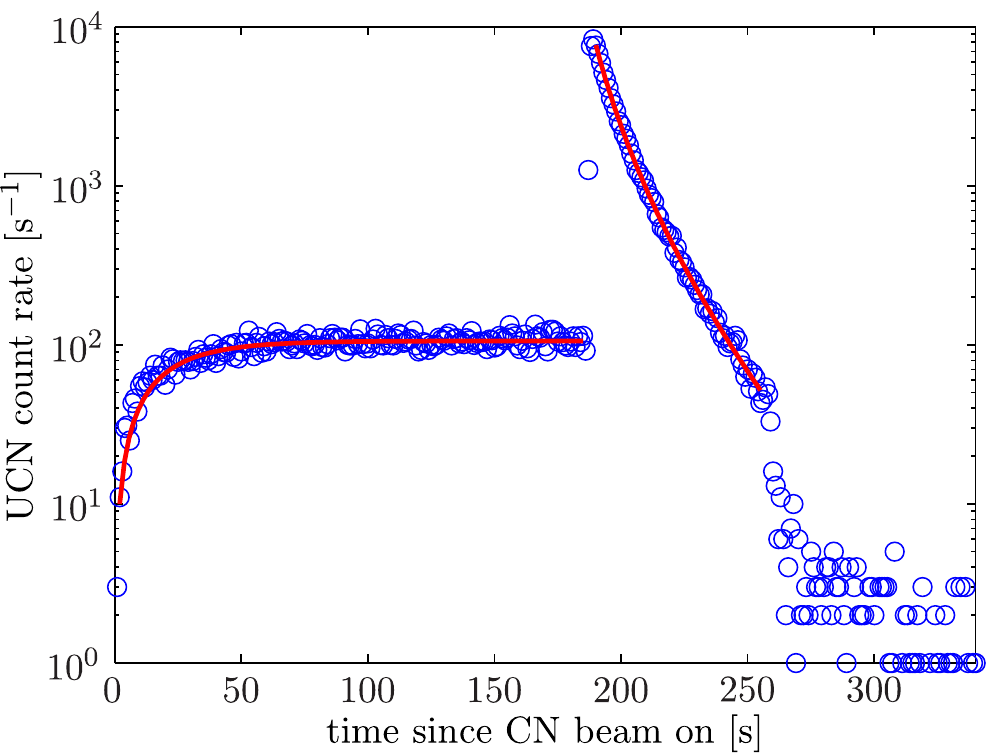}
\caption{(Color online) UCN count rate during the 1.08~K accumulation measurement. At $187\un{s}$ ($\equiv t_0$) the UCN valve was opened and CN beam switched off. At $258\un{s}$ the UCN valve was closed again, before all the UCNs were emptied from the converter volume. The fits of the build-up and emptying curves, details of which are described in the text, are shown as solid lines.}
\label{fig:accumandempty}
\end{center}
\end{figure}

An accumulation measurement with the converter temperature at $T=1.08\un{K}$ was performed. The count rate observed at the UCN detector during this measurement is shown in Fig.~\ref{fig:accumandempty}. Due to poor cooling of the radiation shields surrounding the volume this was the lowest temperature attained in the experiment. 

Due to the UCN valve not being fully tight against leakage of UCNs when closed, UCNs can reach the detector during the accumulation period. The detected rate, denoted by $\dot{N}_{\rm acc}(T,t)$, is related to the number of UCNs in the converter vessel $N(E,T,t)$ from Eq.~\ref{eq:differentialaccumulation}. Since the UCN detector does not provide energy information, it only observes the integral over $E$ for UCNs storable in the vessel. Thus, $\dot{N}_{\rm acc}(T,t)$ is given by (neglecting the time delay due to the time-of-flight of the UCNs)
\begin{eqnarray}
\label{eq:accumulationindetector}
\dot{N}_{\rm acc}(T,t) = \int \epsilon_{\rm ex}(E)\,\tau^{-1}_{\rm leak}(E)\, N_{\infty}(E,T) \,\nonumber \\ \times \left\{ 1 - \exp[-t/\tau_{\rm tot}(E,T)] \right\}\,{\rm d}E\;,
\end{eqnarray}
where $\epsilon_{\rm ex}(E)$ is the efficiency of the UCN extraction and detection, and $\tau^{-1}_{\rm leak}(E)$ is the rate constant of UCNs escaping through the leak at the UCN valve. It will be seen later that $\tau^{-1}_{\rm leak}(E) \ll \tau^{-1}_{\rm tot}(E)$, so the size of leak does not affect $\tau_{\rm tot}(E,T)$. $\dot{N}_{\rm acc}(T,t)$ can be used for monitoring the relative number of UCNs inside the vessel. $\tau^{-1}_{\rm leak}(E)$ and the absolute value of $\dot{N}_{\rm acc}(T,t)$ are not important in this analysis. The limits of integration over $E$ are from the minimum UCN energy that can reach the detector due to vertical extraction to the maximum energy of UCNs storable in the converter and transportable by the guides. 

When the UCN valve is opened and the CN beam switched off, the accumulated UCNs start emptying from the volume. Using Eq.~\ref{eq:differentialexponentialdecay} (similar to Eq.~\ref{eq:accumulationindetector}), the count rate at the detector during emptying ($t>t_0$) is given by:
\begin{eqnarray}
\label{eq:emptyingcurveindetector}
\dot{N}_{\rm empty}(T,t) = \int \epsilon_{\rm ex}(E)\,\tau^{-1}_{\rm v}(E)\, N(E,T,t_0)\,\nonumber \\ \times\exp[-(t-t_0)/\tau_{\rm e}(E,T)]\, dE \;.
\end{eqnarray}

In order to fit the data without \emph{a priori} knowledge of the energy dependence of $
\tau_{\rm tot}(E,T)$ and $\tau_{\rm e}(E,T)$, the integrals of Eqs.~\ref{eq:accumulationindetector} and \ref{eq:emptyingcurveindetector} can be approximated by a discrete sum of terms with different amplitudes $\alpha_i$ and $\beta_i$ and time constants $\tau_{{\rm tot},i}$ and $\tau_{{\rm e},i}$ given by
\begin{equation}
\dot{N}_{\rm acc}(T,t) \approx  \sum_{i=1}^{\eta_{\rm \;\! acc}} \alpha_i \, \left\{1-\exp[-t/\tau_{{\rm tot},i}(T)]\right\}\:
\label{eq:discreteAccumulation}
\end{equation}
and
\begin{equation}
\dot{N}_{\rm empty}(T,t) \approx  \sum_{i=1}^{\eta_{\rm \;\! empty}} \beta_i \,\exp[-(t-t_0)/\tau_{{\rm e},i}(T)]\:.
\label{eq:discreteEmpty}
\end{equation}
The size of $\eta_{\rm \;\! acc}$ or $\eta_{\rm \;\! empty}$ required to satisfactorily describe the data (i.e. $\chi^2_\nu \approx 1$) depends on the strength of the $E$-dependence in $\tau_{\rm tot}(E,T)$ or $\tau_{\rm e}(E,T)$, as well as on the statistical precision of the data.

Using $\eta_{\rm \;\! acc}=1$ is sufficient for the fit of $\dot{N}_{\rm acc}(1.08\un{K},t)$ from $0<t<t_0$ using Eq.~\ref{eq:discreteAccumulation}, which gives $\chi^2_\nu = 0.80$ with $\tau_{\rm tot}(1.08\un{K}) = (20.3 \pm 1.2)\un{s}$. Due to $\eta_{\rm \;\! acc}=1$ being a valid approximation, we drop the $i$-subscript and write simply $\tau_{\rm tot}$.

Using $\eta_{\rm \;\! empty} = 2$ to fit $\dot{N}_{\rm empty}(1.08\un{K},t)$ from $t > t_0$ to when the UCN valve was closed gives $\chi^2_\nu = 1.1$, with the values: $\beta_1 = (5,\!060\pm160)\un{s^{-1}}$, $\tau_{{\rm e},1} = (6.1\pm0.2)\un{s}$, $\beta_2 = (2,\!530\pm180)\un{s^{-1}}$, and $\tau_{{\rm e},2} = (16.3\pm0.4)\un{s}$. For comparison, using $\eta_{\rm \;\! empty} = 1$ gives $\chi^2_\nu = 20$. The need for $\eta_{\rm \;\! empty} > \eta_{\rm \;\! acc}$ is due to the increased statistics of the emptying data and the larger $E$-dependence of $\tau_{\rm e}(E,T)$ (see Eq.~\ref{eq:emptyingtime}).

Following previous papers \cite{Zimmer2010,Zimmer2007a}, we write Eq.~\ref{eq:differentialspectrumequilbrium} in its integral form since it is ensemble averages we are concerned with. The total number of UCNs accumulated in the vessel at saturation can thus be expressed as
\begin{equation}
\label{eq:productionRatetoNumber}
N_\infty(T) =  P(T) \,V_{\rm CN}  \, \tau_{\rm tot}(T)\,.
\end{equation}
This expression can be used to indirectly determine the volumetric UCN production rate $P(T)$.

Summing the counts observed in the emptying curve gives $N_{\rm empty}(1.08\un{K}) = 91,\!700 \pm 300$. To extract $P(T)$, $N_{\rm empty}(T)$ is used to estimate the number of UCNs present inside the converter volume at the end of accumulation, $N(E,T,t_0)$. If we assume that saturated accumulation was reached,  i.e. $N(E,T, t_0) = N_\infty(E,T)$ for all $E$, then the two quantities are related by
\begin{eqnarray}
\label{eq:observedaccumulatedUCNnumber}
N_{\rm empty}(T)  &\equiv&  \int_{t_0}^{\infty}\dot{N}_{\rm empty}(T,t)\,{\rm d}t  \\
&=& \int_{t_0}^{\infty} \epsilon_{\rm ex}(E)\,\epsilon_{\rm v}(E,T)\,N_\infty(E,T)\,{\rm d}E \,, \nonumber
\end{eqnarray}
where
\begin{equation}
\epsilon_{\rm v}(E,T) \equiv \frac{\tau_{\rm e}(E,T)}{\tau_{\rm v}(E)} \approx \frac{\tau_{\rm tot}(E,T) - \tau_{\rm e}(E,T)}{\tau_{\rm tot}(E,T)} \;
\label{eq:definitionepsilonvessel}
\end{equation}
is the probability a UCN can find the valve exit before becoming lost in the vessel.

The discrete $\eta_{\rm \;\! empty} = 2$ approximation of the emptying curve (Eq.~\ref{eq:discreteEmpty}) can be applied to Eq.~\ref{eq:observedaccumulatedUCNnumber} to give
\begin{equation}
\label{eq:discretePopEmptying}
N_{\rm empty}(T) = \epsilon_{\rm ex,1}\,\epsilon_{\rm v,1}\,N_{\infty,1} + \epsilon_{\rm ex,2}\,\epsilon_{\rm v,2} \, N_{\infty,2} \;,
\end{equation}
where $\epsilon_{{\rm v},i} = [\tau_{\rm tot}-\tau_{{\rm e},i}]/\tau_{\rm tot}$, $N_{\rm \infty,i} = \beta_i \tau_{{\rm e},i}/(\epsilon_{{\rm ex},i}\,\epsilon_{{\rm v},i})$, and $N_{\rm \infty} = N_{\rm \infty,1}+N_{\rm \infty,2}$. If we define an effective extraction and detection probability value
\begin{equation}
\bar{\epsilon}_{\rm ex} = \frac{ \epsilon_{\rm ex,1}\,\epsilon_{\rm v,1}\,N_{\infty, 1} + \epsilon_{\rm ex,2}\,\epsilon_{\rm v,2}\,N_{\infty, 2}}{\epsilon_{\rm v,1}\,N_{\infty, 1} + \epsilon_{\rm v,2}\,N_{\infty, 2}}\;,
\end{equation}
then we can write 
\begin{equation}
\label{eq:barepsilonv}
N_{\rm empty}(T) = \bar{\epsilon}_{\rm ex} (\epsilon_{\rm v,1}\,N_{\infty, 1} + \epsilon_{\rm v,2}\,N_{\infty, 2}) \;.
\end{equation}
$\bar{\epsilon}_{\rm ex}$ is the weighted average of $\epsilon_{\rm ex,1}$ and $\epsilon_{\rm ex,2}$ with weights $\epsilon_{\rm v,1}\,N_{\infty, 1}$ and $\epsilon_{\rm v,2}\,N_{\infty, 2}$, respectively. 
Similarly, we can further define
\begin{equation}
\bar{\epsilon}_{\rm v} = \frac{\epsilon_{\rm v,1}\,N_{\infty, 1} + \epsilon_{\rm v,2}\,N_{\infty, 2}}{N_{\infty, 1} + N_{\infty, 2}} \; ,
\end{equation}
so that
\begin{equation}
\label{eq:epexepcN}
N_{\rm empty}(T) = \bar{\epsilon}_{\rm ex}\,\bar{\epsilon}_{\rm v} (N_{\infty, 1} + N_{\infty, 2}) \; .
\end{equation}

From combining Eqs.~\ref{eq:epexepcN} and \ref{eq:observedaccumulatedUCNnumber}, the volumetric UCN production rate, without making corrections for loss mechanisms, can be deduced to be
\begin{equation}
\bar{\epsilon}_{\rm ex}\,\bar{\epsilon}_{\rm v}(1.08\un{K})\,P(1.08\un{K}) = (1.96 \pm 0.14)\un{cm^{-3}\,s^{-1}} \; .
\end{equation}

If we wish to correct for the loss due to $\epsilon_{\rm v}(E,T)$, we put the values for $\epsilon_{{\rm v},i}$ used in Eq.~\ref{eq:discretePopEmptying} into Eq.~\ref{eq:barepsilonv} and then combine with Eq.~\ref{eq:observedaccumulatedUCNnumber} to get
\begin{equation}
\bar{\epsilon}_{\rm ex}\,P(1.08\un{K}) = (6.9 \pm 1.7)\un{cm^{-3}\,s^{-1}} \; .
\end{equation}

The correction for $\bar{\epsilon}_{\rm ex}$ is more difficult to make since it can not be derived from observed quantities in this experiment, and it depends on the UCN spectrum. UCN transport and detection is however not expected to be strongly $E$-dependent due to the few reflections (compared to storage) and the vertical drop before the detector. The detection efficiency and transmission at the window our UCN detector is $\sim80\%$. Typical transmission per unit length of polished stainless steel UCN guides of this diameter are $\sim 90\un{\%/m}$ \cite{Golub1991}. And in a previous experiment with a similar UCN extraction guide geometry and UCN detector of the same type, from comparing with \emph{in situ} CN beam characterization measurements, a value of $\bar{\epsilon}_{\rm ex} \approx 38\%$ was obtained \cite{Zimmer2010}.

The PF1B volumetric UCN production rate was calculated to be $(13.9\pm0.9)\un{cm^{-3}\,s^{-1}}$ in Ref.~\cite{Schmidt-Wellenburg2009b} by using the beam characterization data from Ref.~\cite{Abele2006} and an upper energy cut-off of $U_{\rm Be}-U_{\rm He}$, where $U_{\rm He} = 18.5\un{neV}$ is the optical potential of the converter. This value has no adjustment for vertical extraction.

When UCNs leave the helium, they get a boost in kinetic energy of $U_{\rm He}$. The height of the vertical extraction, measured from the flap valve to the horizontal guide, is $h_{\rm ex} = 28\un{cm}$, which corresponds to a potential difference of $U_{\rm ex} = 29\un{neV}$. Assuming a constant phase space density production spectrum, i.e. $P(E,T) \propto \sqrt{E}$, the correction factor of the production rate from Ref.~\cite{Schmidt-Wellenburg2009b} for vertical extraction is: $ [(U_{\rm Be} - U_{\rm He})^{3/2}- (U_{\rm ex}-U_{\rm He})^{3/2}]/(U_{\rm Be}-U_{\rm He})^{3/2} = 0.96$. This corresponds to a volumetric production rate of $P = (13.3\pm 0.9)\un{cm^{-3}\,s^{-1}}$. 

In our setup the UCN extraction guides are stainless steel so it is not expected that all UCNs with $E$ up to $U_{\rm Be}$ outside the converter make it to the detector. (UCNs with $E>U_{\rm SS}$ can only do so if they only reflect off the UCN guides above the level of the converter, or if they make only glancing reflections off the guides only.) A pessimistic case is that all UCNs with $E>U_{\rm SS}$ outside the converter are lost, then the correction factor becomes: $ [(U_{\rm SS} - U_{\rm He})^{3/2}- (U_{\rm ex}-U_{\rm He})^{3/2}]/(U_{\rm Be}-U_{\rm He})^{3/2} = 0.67$, corresponding to $P = (9.3\pm 0.6)\un{cm^{-3}\,s^{-1}}$. These values do not take into account the variation of the actual CN flux compared to that from Ref.~\cite{Abele2006} as mentioned earlier.

\section{1.2~K to 2.4~K Continuous Measurements}

\begin{figure}[htbp]
\begin{center}
\includegraphics[width=\columnwidth]{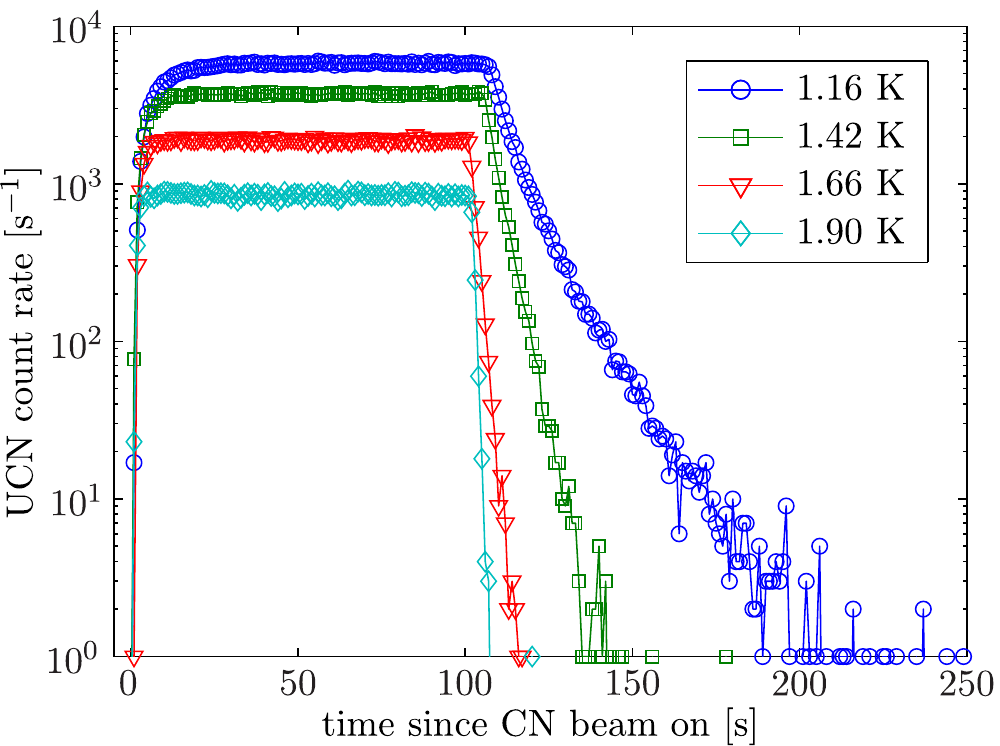}
\caption{(Color online) Continuous measurements for selected temperatures to demonstrate the changes in the build-up, saturation, and emptying. The duration the CN beam is left on is not the same between the measurements. The last 50--60\un{s} before the CN beam is switched off is used to determine $\dot{N}_{\rm c}(T)$.}
\label{fig:continousData}
\end{center}
\end{figure}

\begin{figure}[htbp]
\begin{center}
\includegraphics[width=\columnwidth]{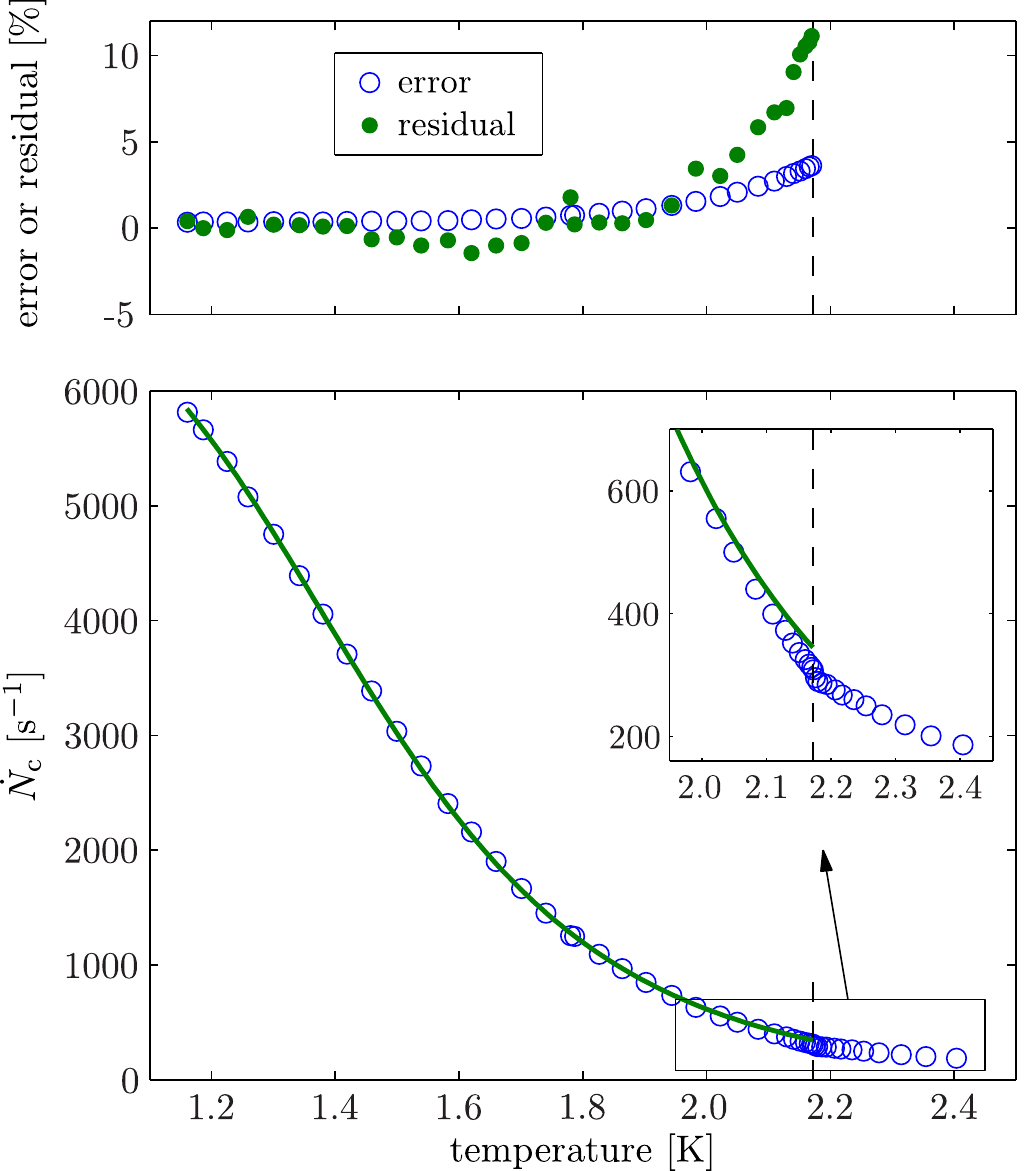}
\caption{(Color online) \emph{Bottom}: The saturated count rates from continuous measurements $\dot{N}_{\rm c}$ between $\unit[1.16\,\text{--}\,2.40]{K}$. The region around $T_\lambda$ is shown zoomed-in. The solid green line is the fit for $T < T_\lambda$ using $\tau_{\rm up}^{-1}(T) = \tau^{-1}_\textrm{2-ph}$. $\dot{N}_{\rm c}(T)/P_{\rm rel}(T)$ is being fitted in the analysis, but $\dot{N}_{\rm c}(T)$ is shown here. See text for details. \emph{Top}: The relative error of each $\dot{N}_{\rm c}(T)/P_{\rm rel}(T)$ point and the residuals (\emph{fit minus data}) of the $\tau_{\rm up}^{-1}(T) = \tau^{-1}_\textrm{2-ph}$ fit are also shown. $T_\lambda$ is marked by the vertical dashed lines.}
\label{fig:continousNc}
\end{center}
\end{figure}

Continuous measurements over a range of converter temperatures were used to study $\tau^{-1}_{\rm up} (T)$. The UCN count rates for selected measurements are shown in Fig.~\ref{fig:continousData}, and the saturated count rates $\dot{N}_{\rm c}(T)$ as a function of temperature are plotted in Fig.~\ref{fig:continousNc}. The temperature during each measurement was stable to $< 10\un{mK}$.

$\dot{N}_{\rm c}(T)$ is given by the rate of continuously produced UCNs in the converter vessel that find the exit of the valve (Eq.~\ref{eq:definitionepsilonvessel}) and get detected; i.e.,
\begin{equation}
\dot{N}_{\rm c}(T) = \int P(E,T) \, V_{\rm CN}\, \epsilon_{\rm ex}(E)\,\frac{\tau_{\rm v}^{-1}(E)}{\tau_{\rm e}^{-1}(E,T)}\, {\rm d}E\;.
\label{eq:Ncdescriptionintegral}
\end{equation}

Since all the temperature-dependence is isolated in $\tau^{-1}_{\rm up}(T)$, an expression appropriate for fitting $\dot{N}_{\rm c}(T)$ to study its temperature-dependence can be written as
\begin{eqnarray}
\label{eq:fittingeqtauecontinuous}
& &\dot{N}_{\rm c}(T) = \frac{P(T) \, V_{\rm CN}\, \epsilon_{\rm ex}\, \tau_{\rm v}^{-1}}{\tau_{0}^{-1}+\tau_{\rm up}^{-1}(T) + \tau_{\rm abs}^{-1} + \tau_{\rm \beta}^{-1} + \tau_{\rm v}^{-1}} \nonumber \\
\Rightarrow  & &\frac{\dot{N}_{\rm c}(T) }{P_{\rm rel}(T)} = \frac{P_0 \, V_{\rm CN}\, \epsilon_{\rm ex} \tau_{\rm v}^{-1}}{\tau_{{\rm e}, T\rightarrow0}^{-1} + \tau_{\rm up}^{-1}(T)}\, ,\\ \nonumber
\end{eqnarray}
which comes from expressing the integral of Eq.~\ref{eq:Ncdescriptionintegral} with the ensemble averaged values $\epsilon_{\rm ex}$, $\tau^{-1}_{\rm v}$, and $\tau^{-1}_{0}$. We do not require to know these values, they need only to be constant with temperature. All the rate constants contributing to $\tau_{\rm e}^{-1}$ besides $\tau_{\rm up}^{-1}(T)$ are combined into $\tau_{{\rm e},T\rightarrow0}^{-1}$, which is used as one of the two free parameters of the data fitting. The other free parameter is the scaling factor $P_0 \, V_{\rm CN}\, \epsilon_{\rm ex}\, \tau_{\rm v}^{-1}$, where $P_0 \equiv P(T\rightarrow0\un{K})$. The $T$-dependence of production is contained in the relative UCN production rate $P_{\rm rel}(T)$, i.e. $P_{\rm rel}(T) = P(T)/P_0$.

$P_{\rm rel}(T)$, required for the data-fitting, is calculated via the van Hove relation \cite{Korobkina2002} using experimental data from Ref.~\cite{Andersen1994a} as described in the Appendix. The results are shown in Fig.~\ref{fig:relativeUCNproductionRate}. $P_{\rm rel}(T)$ decreases with temperature as $T_\lambda = 2.172\un{K}$ is approached and then increases. Qualitatively, the kink in the $\dot{N}_{\rm c}(T)$ data at $T_\lambda$ can be explained by $P_{\rm rel}(T)$. Accurately describing $\dot{N}_{\rm c}(T > T_\lambda)$ is however difficult due to effects such as helium vapor bubbling in the liquid, and is not attempted here.

\begin{figure}[htbp]
\begin{center}
\includegraphics[width=\columnwidth]{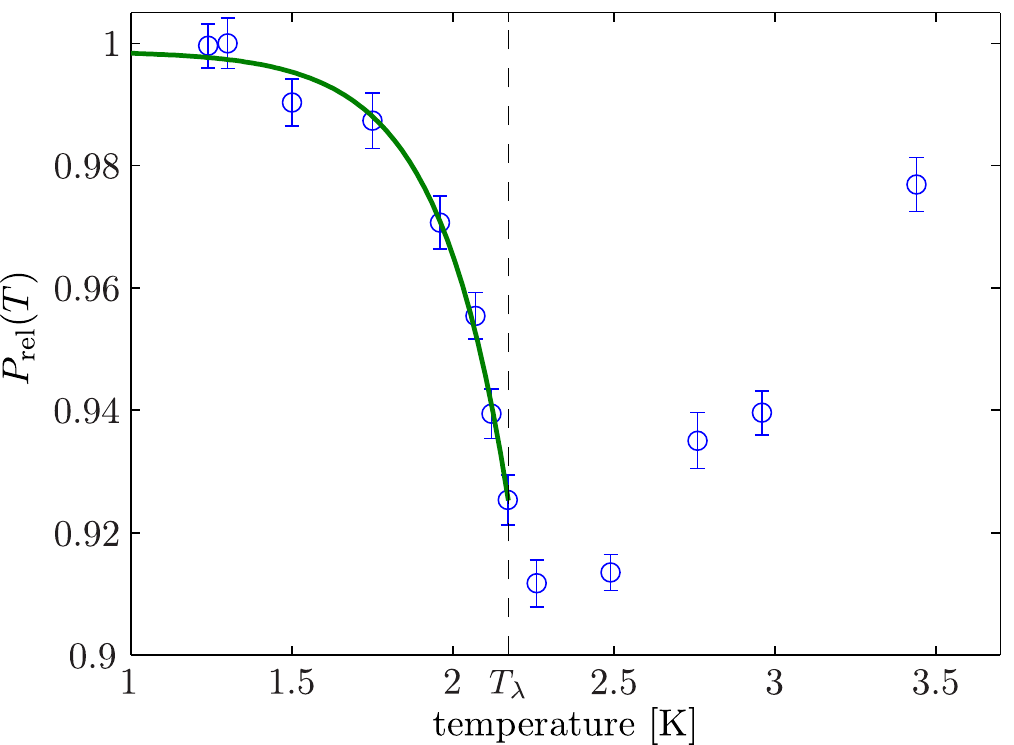}
\caption{(Color online) The relative UCN production rate $P_{\rm rel}(T)$ calculated using inelastic neutron scattering data and an analytic approximation of the PF1B CN beam spectrum (see the Appendix for details of the calculations). The solid green line is the interpolation between the points with Eq.~\ref{eq:UCNproductionApproximation} used in the data analysis.}
\label{fig:relativeUCNproductionRate}
\end{center}
\end{figure}

For the data analysis, interpolation between the $P_{\rm rel}(T)$ values is required. Below $T_\lambda$, with no known physical significance, the expression
\begin{equation}
\label{eq:UCNproductionApproximation}
P_{\rm rel}(T<T_\lambda) \approx p_1\left\{ 1-\exp[(T-p_2)/p_3]\right\}\;
\end{equation}
is used to fit the $P_{\rm rel}(T)$ values for this purpose. This gives $p_1 = 0.999\pm0.003$, $p_2 = (2.74\pm 0.10)\un{K}$, and $p_3 = (0.22 \pm 0.04)\un{K}$, with a $\chi^2_{\nu} = 0.62$.

The uncertainty in the $\dot{N}_{\rm c}(T)/P_{\rm rel}(T)$ points used for fitting with Eq.~\ref{eq:fittingeqtauecontinuous} comes from combining in quadrature the statistical uncertainty in $\dot{N}_{\rm c}(T)$ ($0.2\,\text{--}\,0.8\%$) and the uncertainty from the interpolation of $P_{\rm rel}(T)$ ($0.3\,\text{--}\,3\%$). This is shown in Fig.~\ref{fig:continousNc}.

To determine the best form of $\tau^{-1}_{\rm up}$ that describes the data, non-linear fitting with different combinations of the terms for $\tau_{\rm up}^{-1}(T)$ from Eq.~\ref{eq:tauuptheoretical} is done. A comparison between fitting with the fixed coefficients $A = 130\un{s^{-1}}$, $B =  7.6 \times 10^{-3} \un{s^{-1}\,K^{-7}}$, and $C = 18\un{s^{-1}\, K^{-3/2}}$ for $\tau_\textrm{1-ph}^{-1}$, $\tau_\textrm{2-ph}^{-1}$, and $\tau_\textrm{rot-ph}^{-1}$, respectively, is shown in Table~\ref{tab:ABCfixed}. In all the analyses, the starting values of the optimization routine was varied to ensure we are finding the global $\chi^2_\nu$ minimum.

When $T<T_\lambda$ is used the best fit is obtained using $\tau_{\rm up}^{-1}(T) = \tau_\textrm{2-ph}^{-1}$ producing $\chi^2_\nu = 3.8$. A plot of this fit and its residual is shown in Fig.~\ref{fig:continousNc}. The next best fit comes from using $\tau_{\rm up}^{-1}(T) = \tau_\textrm{1-ph}^{-1} + \tau_\textrm{2-ph}^{-1}$ with $\chi^2_\nu = 8.1$. The former fit, according to the $\chi^2_\nu$ distribution ($\nu = 30$) \cite{Bevington2003} is $10^{23}$ times more likely to be a better model for the data. Any fit that includes $\tau_\textrm{rot-ph}^{-1}$, which should be comparable in size at this temperature range, is much poorer.

\begin{table}[htbp]
\begin{center}
\caption{Fitting $\dot{N}_{\rm c}(T<T_\lambda)$ and $\dot{N}_{\rm c}(T<1.95\un{K})$ with Eq.~\ref{eq:fittingeqtauecontinuous} for different combinations of $\tau_{\rm up}^{-1}$ and using  fixed coefficients $A$, $B$, and $C$ according to theoretical predictions (see discussion around Eq.~\ref{eq:tauuptheoretical}). The best-fit parameters and reduced chi-squared $\chi^2_\nu$ ($\nu  = 30 \text{ and } 19$ respectively) are tabulated. Bold font indicates the best fit.}
\begin{tabular}{|c|c|c|c|}
\hline
\multirow{2}{*}{\normalsize $\tau^{-1}_{\rm up}$} & {\footnotesize $P_0V_{\rm CN}\epsilon_{\rm ex}\tau_{\rm v}^{-1}$} & \multirow{2}{*}{\normalsize $\tau_{{\rm e},T\rightarrow0}^{-1}$ [s]}  & \multirow{2}{*}{{\small $\chi^2_\nu$}} \\ 
&{\small {\footnotesize [s$^{-2}$]}} 	&  & \\ \hline
\multicolumn{4}{ |c| }{{\small  $T<T_\lambda$ ($\nu = 30$)}} \\  \hline
{\small $\tau_\textrm{1-ph}^{-1}$} & 245$\,\pm\,$1 & 26.25$\,\pm\,$0.10 & 31 \\ \hline
{\small ${\tau_\textrm{2-ph}^{-1}}$} & {\bf 680$\,\pm\,$2} & {\bf 10.56$\,\pm\,$0.05} & $\mathbf{3.8}$  \\ \hline
{\small $\tau_\textrm{rot-ph}^{-1}$} & 547$\,\pm\,$1 & 12.77$\,\pm\,$0.05 & 56  \\ \hline
{\footnotesize $\tau_\textrm{2-ph}^{-1} + \tau_\textrm{rot-ph}^{-1} $} & 1229$\,\pm\,$3 & 5.77$\,\pm\,$0.02 & 18  \\ \hline
{\footnotesize $\tau_\textrm{1-ph}^{-1} + \tau_\textrm{2-ph}^{-1} $} & 927$\,\pm\,$2 & 7.52$\,\pm\,$0.03 & 8.1  \\ \hline
{\footnotesize $\tau_\textrm{1-ph}^{-1} + \tau_\textrm{rot-ph}^{-1} $} & 792$\,\pm\,$2 & 8.59$\,\pm\,$0.04 & 46  \\ \hline
{\tiny $\tau_\textrm{1-ph}^{-1} + \tau_\textrm{2-ph}^{-1} + \tau_\textrm{rot-ph}^{-1} $} & 1475$\,\pm\,$4 & 4.73$\,\pm\,$0.02 & 19  \\ \hline
\multicolumn{4}{ |c| }{\small  $T<1.95\un{K}$  ($\nu = 19$) } \\  \hline
{\small $\tau_\textrm{1-ph}^{-1}$} & 251$\,\pm\,$1 & 25.36$\,\pm\,$0.11 & 11 \\ \hline
{\small ${\tau_\textrm{2-ph}^{-1}}$} & {\bf 686$\,\pm\,$2} & {\bf 10.44$\,\pm\,$0.05} & $\mathbf{1.7}$  \\ \hline
{\small $\tau_\textrm{rot-ph}^{-1}$} & 564$\,\pm\,$1 & 12.19$\,\pm\,$0.05 & 25  \\ \hline
{\footnotesize $\tau_\textrm{2-ph}^{-1} + \tau_\textrm{rot-ph}^{-1} $} & 1250$\,\pm\,$3 & 5.62$\,\pm\,$0.03 & 8.0  \\ \hline
{\footnotesize $\tau_\textrm{1-ph}^{-1} + \tau_\textrm{2-ph}^{-1} $} & 937$\,\pm\,$2 & 7.39$\,\pm\,$0.03 & 3.0  \\ \hline
{\footnotesize $\tau_\textrm{1-ph}^{-1} + \tau_\textrm{rot-ph}^{-1} $} & 815$\,\pm\,$2 & 8.24$\,\pm\,$0.04 & 18  \\ \hline
{\tiny $\tau_\textrm{1-ph}^{-1} + \tau_\textrm{2-ph}^{-1} + \tau_\textrm{rot-ph}^{-1} $} & 1501$\,\pm\,$4 & 4.60$\,\pm\,$0.02 & 7.7  \\ \hline

\end{tabular}
\label{tab:ABCfixed}
\end{center}
\end{table}

As can be seen from the fit residual shown in Fig.~\ref{fig:continousNc}, a large contribution to the $\chi^2_\nu$ comes from $T > 1.95\un{K}$. This is true for the other fits as well. At these higher temperatures, approximations used in the fit break down on the $\sim 10\%$ level. This indicates there are new phenomena causing either an increase in $\tau^{-1}_{\rm up}(T)$ or $\tau^{-1}_{\rm v}$, or reducing $P_{\rm rel}(T)$ beyond that predicted by the calculations with neutron scattering data. Since these new effects are not included in the description provided thus far, it is more valid to restrict the data fitting to $T<1.95\un{K}$ only. The results of repeating the previous analysis in this temperature range is also shown in Table~\ref{tab:ABCfixed}. The same conclusion is reached: the best fit is still from $\tau_{\rm up}^{-1} = \tau_\textrm{2-ph}^{-1}$, with the $\chi^2_\nu$ decreasing to 1.7, while the rest of the $\chi^2_\nu$ values decrease by a factor of $\sim 3$. The probability that $\tau_{\rm up}^{-1}(T) = \tau_\textrm{2-ph}^{-1}$ is a more likely fit than $\tau_{\rm up}^{-1}(T) = \tau_\textrm{1-ph}^{-1} + \tau_\textrm{2-ph}^{-1}$ is now $3000$ times greater only  \cite{Bevington2003}. In all subsequent analyses, we restrict to using $T<1.95\un{K}$.

Also explored are potentially incorrect fixed values for $A$, $B$, and $C$ affecting the quality of the fits. To do this, the values were varied to see their impact on the $\chi^2_\nu$. This is done by allowing these values to be optimized by the Levenberg-Marquardt non-linear regression data-fitting algorithm used in the analysis so far. However, if the $\chi^2_\nu$ dependence of a certain variable is weak, then the algorithm does not converge. In this case, the variable was varied manually.

First, we make the fit using $\tau_{\rm up}^{-1}(T) = \tau_\textrm{2-ph}^{-1}$ and vary the parameter $B$. This had to be done manually since the $\chi^2_\nu$ does not change for a wide-range of $B$ values, as shown in Table~\ref{tab:Bvaried}. $\chi^2_\nu$ stays the same because the values of $\tau_{{\rm e},T\rightarrow0}$ and $P_0V_{\rm CN}\epsilon_{\rm ex}\tau_{\rm v}^{-1}$ are able to compensate, however, at some stage $\tau_{{\rm e},T\rightarrow0}$ becomes unphysical when compared to the observed $\tau_{{\rm e},i}(1.08\un{K})$ values from the accumulation measurement of $(6.1\pm0.2)\un{s}$ and $(16.3\pm0.4)\un{s}$ (see previous section). These values are expected to be similar to those when $T\rightarrow 0$. We thus make the assertion that $\tau_{{\rm e},T\rightarrow0}$ should fall in the range conservative range of $(5\text{ -- }20)\un{s}$. This leads to a value for $B =(4\text{ -- } 16) \times 10^{-3}\un{s^{-1}\,K^{-7}}$ being sufficient to describe the data. The fitted $P_0V_{\rm CN}\epsilon_{\rm ex}\tau_{\rm v}^{-1}$ values can be compared to the expected value too, however, this is much less stringent due to the number of parameters required to be estimated.

\begin{table}[htbp]
\begin{center}
\caption{Fitting $\dot{N}_{\rm c}(T<1.95\un{K})$ with Eq.~\ref{eq:fittingeqtauecontinuous} using $\tau_{\rm up}^{-1}(T) = \tau_\textrm{2-ph}^{-1}$ and with $B$ varied manually. Bold font indicates fits that satisfy the acceptable limits for $\tau_{{\rm e},T\rightarrow0}$ (see text).}
\begin{tabular}{|c|c|c|c|c|}
\hline
{\small $B$} & \multirow{2}{*}{\emph{0.76}} & \multirow{2}{*}{\emph{4}} & \multirow{2}{*}{\emph{16}} & \multirow{2}{*}{\emph{76}}\\
{\footnotesize [$10^{\text{-}3}\;{\rm s^{\text{-}1}\,K^{\text{-}7}}$]} & & & & \\ \hline

$\chi^2_\nu$ & \small 1.7 & \small \bf 1.7  & \small \bf 1.7 & \small 1.7 \\\hline

{\footnotesize $P_0V_{\rm CN}\epsilon_{\rm ex}\tau_{\rm v}^{-1}$} &\multirow{2}{*}{\footnotesize 69$\,\pm\,$0.2} &\multirow{2}{*}{\bf\footnotesize 361$\,\pm\,$1} &\multirow{2}{*}{\bf\footnotesize 1443$\,\pm\,$4} & \multirow{2}{*}{\footnotesize 6860$\,\pm\,$20} \\
{\small [${\rm s^{-2}}$]} & & & &\\ \hline

{\normalsize $\tau_{{\rm e},T\rightarrow0}$ [s]} &{\footnotesize 104$\,\pm\,$0.5} &{\bf\footnotesize 19.8$\,\pm\,$0.1}  &{\bf\footnotesize 4.95$\,\pm\,$0.02} &{\footnotesize 1.044$\,\pm\,$0.005} \\\hline

\end{tabular}
\label{tab:Bvaried}
\end{center}
\end{table}

Next we analyze the effect of varied coefficients when $\tau_{\rm up}^{-1}(T) = \tau_\textrm{2-ph}^{-1} + \tau_\textrm{rot-ph}^{-1}$ is used for the fit. Like previously, $B$ has to be varied manually, however $C$ can be optimized with the fitting algorithm. The results of this analysis are shown in Table~\ref{tab:BCvaried}. In the optimization, the values of $C$ are restricted to $C\geq 0$. It can be seen that the optimized $C$ values for all values of $B$ used are consistent with zero. Because of this, the optimized $\tau_{{\rm e},T\rightarrow0}$ and $\chi^2_\nu$ are similar to those from the fits with $\tau_{\rm up}^{-1}(T) = \tau_\textrm{2-ph}^{-1}$ and $B$ fixed. The weak dependence of the fit on $B$, and the previously quoted acceptable range for $B$, are again confirmed. From this analysis, we deduce the largest possible value for $C$ is $<5.3\un{s^{-1}\,K^{-3/2}}$ (95\% C.I.) in order to keep $\tau_{{\rm e},T\rightarrow0} \sim (5\text{ -- }20)\un{s}$, which equates to $B =  (4\text{ -- } 16) \times 10^{-3}\un{s^{-1}\,K^{-7}}$, the same range from the previous analysis.

\begin{table*}[htbp]
\begin{center}
\caption{Fitting $\dot{N}_{\rm c}(T<1.95\un{K})$ using Eq.~\ref{eq:fittingeqtauecontinuous} and $\tau_{\rm up}^{-1}(T) = \tau_\textrm{2-ph}^{-1} + \tau_\textrm{rot-ph}^{-1}$ with $B$ varied manually and $C$ optimized algorithmically. Bold font indicates solutions that satisfy the acceptable limits for $\tau_{{\rm e},T\rightarrow0}$ (see text).}
\begin{tabular}{|c|c|c|c|c|c|}
\hline
$B$ {\normalsize [$10^{-3}\;{\rm s^{-1}\,K^{-7}}$]} & \emph{0.76} & \emph{4} & \emph{7.6} &  \emph{16} & \emph{76}\\\hline
$\chi^2_\nu$ & 1.8 & \bf 1.8 & \bf 1.8 & \bf 1.8 & 1.8 \\\hline
$C$ {\normalsize [${\rm s^{-1}\, K^{-3/2}}$]} &{\normalsize $(1\times10^{-4})\pm0.1$} &{\normalsize $\bf(1\times 10^{-3}) \pm 0.7$} &{\normalsize $\bf(3 \times10^{-4}) \pm 1.3$} &{\normalsize $\bf(4\times10^{-3})\pm 2.7$} &{\normalsize $(2\times10^{-2}) \pm 1.3$} \\\hline
{\normalsize $\tau_{ {\rm e},T\rightarrow0}$ [s]} &{\normalsize 104$\,\pm\,$6} &{\bf \normalsize 20$\,\pm\,$1} &{\bf\normalsize 10.4$\,\pm\,$0.6} &{\bf\normalsize 5.0$\,\pm\,$0.3} &{\normalsize 1.04$\,\pm\,$0.06} \\\hline
\end{tabular}
\label{tab:BCvaried}
\end{center}
\end{table*}

Finally, we vary all three coefficients in the $\tau_{\rm up}^{-1}(T) = \tau_\textrm{1-ph}^{-1} + \tau_\textrm{2-ph}^{-1} + \tau_\textrm{rot-ph}^{-1}$ fit. The value of $A$ also has to be varied manually; as before, $B$ has to be varied manually; and $C$ can be optimized algorithmically. The results of this analysis are shown in Table~\ref{tab:ABCvaried}. Both $A$ and $B$ have been varied by two orders of magnitude or more but all the optimized $C$ values are still consistent with zero. For all values of $B$, the smaller the $A$ value the lower the $\chi^2_\nu$. This further reinforces the absence of $\tau_\textrm{rot-ph}^{-1}$ in the data, as well as demonstrates the absence of $\tau_\textrm{1-ph}^{-1}$.

Since this analysis is the joint variation of two parameters ($A$ and $B$) with the other 3 parameters ($\tau_{{\rm e},T\rightarrow0}$, $P_0V_{\rm CN}\epsilon_{\rm ex}\tau_{\rm v}^{-1}$, and $C$) optimized by the Levenberg-Marquardt algorithm, we have separated the $\chi^2$ fitting problem into two parts: a variation of $\chi^2$ for the fit of the data to 3 parameters (with $\nu = 18$) and a variation of $\Delta \chi^2$ about the minimum $\chi^2$ with 2 parameters ($\nu=2$). These two variations separately follow their appropriate $\chi^2$-distributions \cite{Bevington2003}. The limits of 95\% C.I. are thus given by contours of $\text{min}(\chi^2) + 6.14$ (or $\text{min}(\chi^2_\nu) +0.34 = 2.17$).

The limit of $A$ is found by finding the widest part of the 95\% C.I. contour that also satisfies the restriction that $\tau_{{\rm e},T\rightarrow0} = (5\text{ -- }20)\un{s}$, which again equates to $B = (4\text{ -- } 16) \times 10^{-3}\un{s^{-1}\,K^{-7}}$. This occurs for $B = 14 \times 10^{-3}\un{s^{-1}\,K^{-7}}$, and sets the limit $A < 61\un{s^{-1}}$ (95\% C.I.). The limit on $C$ requires having $A$ optimized for each fit \cite{Bevington2003}, which occurs for $A \rightarrow 0$. This provides the limit of $C<5.3\un{s^{-1}\,K^{-3/2}}$ (95\% C.I.) from this analysis, the same as the limit from the previous analysis. 

\begin{table*}[htbp]
\caption{Fitting $\dot{N}_{\rm c}(T<1.95\un{K})$ with Eq.~\ref{eq:fittingeqtauecontinuous} using $\tau_{\rm up}^{-1}(T) = \tau_\textrm{1-ph}^{-1} + \tau_\textrm{2-ph}^{-1} + \tau_\textrm{rot-ph}^{-1}$ with $A$ varied manually, $B$ varied manually, and $C$ optimized algorithmically. Bold font indicates solutions that both satisfy the acceptable limits for $\tau_{{\rm e},T\rightarrow0}$ as well as the 95\% C.I. limit of $\chi^2_\nu$ (see text).}
\begin{tabular}{c c c | c|c|c|c|c|}
\cline{4-8}
& & & \multicolumn{5}{ c| }{\hspace{0cm} $A$ [${\rm s^{-1}}$]} \\ \cline{4-8}	

& &  & \emph{1.3}  & \emph{18}&  \emph{33} &  \emph{61} & \emph{130}  \\ \cline{1-8}
\multicolumn{1}{|c}{\multirow{1}{*}{\rotatebox{90}{$B$ [$10^{-3}\;{\rm s^{-1}\,K^{-7}}$]\hspace{1.5cm}}}}  & \multicolumn{1}{|c|}{}  &$\chi^2_\nu$       & 2.0 & 3.7 & 4.8 & 6.3 & 8.1   \\ 

\multicolumn{1}{|c}{} &\multicolumn{1}{|c|}{\emph{0.76}}& $C$ $ [{\rm s^{-1}\, K^{-3/2}}]$  &{\normalsize $(1\times 10^{-4})\pm 0.1$} &{\normalsize $(3\times 10^{-5})\pm0.2$} &{\normalsize $(6\times 10^{-5})\pm 1.4$}  & {\normalsize $(5\times 10^{-5})\pm0.4$} &{\normalsize $(1\times 10^{-5})\pm0.8$} \\

\multicolumn{1}{|c}{} & \multicolumn{1}{|c|}{} & $\tau_{ {\rm e},T\rightarrow0}$ [s] &{\normalsize 100$\,\pm\,$6}   &{\normalsize 66$\,\pm\,$4}  &{\normalsize  51$\,\pm\,$3}  & {\normalsize 36$\,\pm\,$2} &{\normalsize 20$\,\pm\,$1} \\ \cline{2-8}
\multicolumn{1}{|c}{} &\multicolumn{1}{|c|}{} &  $\chi^2_\nu$ & \bf 1.9 & \bf 2.2 & 2.5 & 3.0 & 4.2   \\ 

\multicolumn{1}{|c}{} &\multicolumn{1}{|c|}{\emph{4}}  &  $C$ $ [{\rm s^{-1}\, K^{-3/2}}]$ &{\normalsize $\bf(6\times10^{-4})\pm 0.7$} &{\normalsize $\bf(1\times10^{-3})\pm 0.7$}  &{\normalsize $(4\times10^{-4})\pm 0.8$} & {\normalsize $(4\times10^{-4})\pm 0.9$}  &{\normalsize $(4\times10^{-4})\pm 1.2$}   \\

\multicolumn{1}{|c}{} &\multicolumn{1}{|c|}{} & $\tau_{{\rm e},T\rightarrow0}$ [s]&{\bf\normalsize 20$\,\pm\,$1}   &{\bf\normalsize 18$\,\pm\,$1} &{\normalsize 17$\,\pm\,$1} & {\normalsize 14.5$\,\pm\,$0.9} &{\normalsize 11.1$\,\pm\,$0.7} \\ \cline{2-8}
\multicolumn{1}{|c}{} & \multicolumn{1}{|c|}{}        & $\chi^2_\nu$ 	& \bf 1.8 & \bf 2.0 & \bf 2.2 & 2.5 & 3.2 \\

\multicolumn{1}{|c}{} &\multicolumn{1}{|c|}{\emph{7.6}} & $C$ $ [{\rm s^{-1}\, K^{-3/2}}]$ &{\normalsize $\bf(4\times 10^{-4})\pm 1.3$} &{\normalsize $\bf(2\times 10^{-3})\pm 1.4$}	&{\normalsize $\bf(2\times 10^{-3})\pm 1.4$}	& {\normalsize $(7\times 10^{-4})\pm 1.5$}  &{\normalsize $(3\times 10^{-3})\pm 1.8$}  \\

\multicolumn{1}{|c}{} & \multicolumn{1}{|c|}{} & $\tau_{{\rm e},T\rightarrow0}$ [s] &{\bf\normalsize 10.4$\,\pm\,$0.6} &{\bf\normalsize 9.9$\,\pm\,$0.6} &{\bf\normalsize 9.4$\,\pm\,$0.6} & {\normalsize 8.7$\,\pm\,$0.5} &{\normalsize 7.4$\,\pm\,$0.5}  \\ \cline{2-8}
\multicolumn{1}{|c}{} & \multicolumn{1}{|c|}{}        & $\chi^2_\nu$ 	& \bf 1.8 & \bf 1.9 & \bf 2.0 & \bf2.2 & 2.6 \\

\multicolumn{1}{|c}{} &\multicolumn{1}{|c|}{\emph{14}} & $C$ $ [{\rm s^{-1}\, K^{-3/2}}]$ &{\normalsize $\bf(4\times 10^{-3})\pm 2.3$} &{\normalsize $\bf(4\times 10^{-3})\pm 2.4$}	&{\normalsize $\bf(3\times 10^{-3})\pm 2.5$}	 & {\normalsize $\bf(2\times 10^{-3})\pm 2.6$}  &{\normalsize $(4\times 10^{-3})\pm 2.9$}   \\

\multicolumn{1}{|c}{} & \multicolumn{1}{|c|}{} & $\tau_{{\rm e},T\rightarrow0}$ [s] &{\bf\normalsize 5.7$\,\pm\,$0.3} &{\bf\normalsize 5.5$\,\pm\,$0.3} &{\bf\normalsize 5.4$\,\pm\,$0.3} & {\bf\normalsize 5.1$\,\pm\,$0.3} &{\normalsize 4.6$\,\pm\,$0.3} \\ \cline{2-8}

\multicolumn{1}{|c}{} & \multicolumn{1}{|c|}{}      & $\chi^2_\nu$   & \bf 1.8 &  1.9 &  2.0 & 2.1 & 2.5 \\

\multicolumn{1}{|c}{} &\multicolumn{1}{|c|}{\emph{16}}    &    $C$ $ [{\rm s^{-1}\, K^{-3/2}}]$ &{\normalsize $\bf(5\times 10^{-3})\pm 2.7$} &{\normalsize $(5\times 10^{-3})\pm 2.8$} &{\normalsize $ (8\times 10^{-4})\pm 2.8$} & {\normalsize $(4\times 10^{-3})\pm 2.9$} &{\normalsize $(6\times 10^{-3})\pm 3.2$}  \\

\multicolumn{1}{|c}{} & \multicolumn{1}{|c|}{} & $\tau_{{\rm e},T\rightarrow0}$ [s] &{\bf\normalsize 4.9$\,\pm\,$0.3}   &{ \normalsize 4.8$\,\pm\,$0.3}  &{  \normalsize 4.7$\,\pm\,$0.3}  & {\normalsize 4.5$\,\pm\,$0.3} &{\normalsize 4.1$\,\pm\,$0.3} \\ \cline{2-8}
\multicolumn{1}{|c}{} &\multicolumn{1}{|c|}{}      &  $\chi^2_\nu$ & 1.8 & 1.9 & 1.9 & 1.9 & 2.0 \\

\multicolumn{1}{|c}{} &\multicolumn{1}{|c|}{\emph{76}}   &   $C$ $ [{\rm s^{-1}\, K^{-3/2}}]$ &{\normalsize $(2\times 10^{-2})\pm 13$}  &{\normalsize $(2\times 10^{-2})\pm 13$} &{\normalsize $(2\times 10^{-2})\pm 13$} & {\normalsize $(2\times 10^{-2})\pm 13$} &{\normalsize $(2\times 10^{-2})\pm 13$} \\

\multicolumn{1}{|c}{} &\multicolumn{1}{|c|}{}  & $\tau_{{\rm e},T\rightarrow0}$ [s] &{\normalsize 1.04$\,\pm\,$0.06}  &{\normalsize 1.03$\,\pm\,$0.06}  &{\normalsize 1.03$\,\pm\,$0.06}  & {\normalsize 1.02$\,\pm\,$0.06}  &{\normalsize 1.00$\,\pm\,$0.06} \\ \cline{1-8}
         
\end{tabular}
\label{tab:ABCvaried}
\end{table*}

\section{Conclusion}

This paper presented UCN production measurements using a superfluid $^4$He UCN source with a 4-liter beryllium-coated vessel, mechanical UCN accumulation valve, and vertical, window-less stainless steel extraction system. Experiments were performed at the PF1B CN beam at the ILL.

In the first part, with an accumulation measurement at 1.08\un{K} an uncorrected UCN production rate of $\bar{\epsilon}_{\rm ex}\,\bar{\epsilon}_{\rm v}(1.08\un{K}) P(1.08\un{K}) = (1.96 \pm 0.14)\un{cm^{-3}\,s^{-1}}$ was observed. After correcting for $\bar{\epsilon}_{\rm v}(1.08\un{K})$, whose value is dominated by a poor $\tau_{\rm tot}(T)$, a production rate of $\bar{\epsilon}_{\rm ex} P(1.08\un{K}) = (6.9 \pm 1.7)\un{cm^{-3}\,s^{-1}}$ was extracted. The approximate size of $\bar{\epsilon}_{\rm ex}$ estimated from other experiments is $\sim 40\%$. The calculated production rate is $P\sim (9\text{\,--\,}13)\un{cm^{-3}\,s^{-1}}$, where the range comes from details in the UCN extraction process. Furthermore, these rates are based on older beam characterization measurements \cite{Abele2006} and could be different to the actual flux in the experiment by $\sim 30\%$.

The short $\tau_{\rm tot}(1.08\un{K}) = (20.3 \pm 1.2)\un{s}$ might be caused by a combined effect of UCN losses due to imperfections of the beryllium coatings of the walls and patches with high losses on the complex UCN valve mechanism, which both can be improved. Better cooling of the converter will improve things further, since $\tau_{\rm up}(1.08\un{K})\approx 50\un{s}$. This would increase the density (without corrections) in the converter from the currently observed value, $\bar{\epsilon}_{\rm ex}\,\bar{\epsilon}_{\rm v}(1.08\un{K})\,P(1.08\un{K})\,V_{\rm CN}\,\tau_{\rm tot} /V_{\rm conv}= (23\pm 3)\un{cm^{-3}}$.

For instance, if $\tau_{\rm tot} = 100\un{s}$ were attained, a modest goal for the apparatus (which would make $\bar{\epsilon}_{\rm v}\approx 0.9$), and if the CN beam were better optimized for the converter vessel ($V_{\rm CN}/V_{\rm conv} \approx 1$), then a uncorrected UCN density of $\bar{\epsilon}_{\rm ex}\,\bar{\epsilon}_{\rm v}\,P\,\tau_{\rm tot} \sim 600\un{cm^{-3}}$ would be expected.

In the second part of this paper, continuous measurements with the converter temperature scanned from 1.2 to 2.4~K were used to study UCN production and up-scattering. The three processes expected to contribute to the up-scattering of UCNs $\tau^{-1}_{\rm up}(T)$ are: 1-phonon absorption, 2-phonon scattering, and roton-phonon scattering.

The continuous measurements for $T < 1.95\un{K}$ were fitted using the theoretical temperature-dependences of these processes, with both calculated values of the $A$, $B$, and $C$ coefficients associated with the above processes respectively, as well as by varying them. Our analysis revealed the need to include only 2-phonon scattering, i.e. $\tau^{-1}_{\rm up}(T) = B\,T^7$, with a value of $B = (4\text{\,--\,}16)\times 10^{-3}\un{s^{-1}\,K^{-7}}$ being valid in describing the data. This conclusion is the same as from earlier experiments \cite{Yoshiki1992,Baker2003, Piegsa2014} but performed using an enlarged temperature range and a more thorough data-fitting procedure. Extracted from the analyses are the limits: $C < 5.3\un{s^{-1}\,K^{-3/2}}$ (95\% C.I.), i.e. $< 29\un{\%}$ of its theoretical value; and $A < 61\un{s^{-1}}$ (95\% C.I.), i.e. $< 47\un{\%}$ of its theoretical value.

The temperature-dependence of the UCN production rate, $P_{\rm rel}(T)$, was calculated from inelastic neutron scattering in this paper. The calculated form of $P_{\rm rel}(T)$ qualitatively explains the interesting kink observed in $\dot{N}_{\rm c}(T)$ at $T_\lambda$, and is required for obtaining a good fit of $\dot{N}_{\rm c}(T< 1.95\un{K})$. A discrepancy caused by an increased UCN loss and/or reduction in UCN production was observed when $T>1.95\un{K}$. This is most likely due to us not taking into full account the  details of superfluid helium close to $T_{\lambda}$.

This study demonstrates that there is still a lack of full understanding of the interactions between UCNs and superfluid $^4$He. It has been suggested \cite{Golub1996} that the observed lack of UCN loss due to roton-phonon scattering is a confirmation of the proposed strong preference for co-linear momentum vectors in the three quasi-particle coupling constant when the momentum and energy transfers are large ($> 3.4\un{\text{\AA}^{-1}}$ and $> 1.1\un{meV}$ respectively) \cite{Halley1991}. Further theoretical work, such as that done in Ref.~\cite{Faak2012}, but for the neutron, could shed light on the current discrepancy. Also, expanding on experimental work, such as improving the precision and temperature range of measurements of the up-scattered UCN spectrum \cite{Gutsmiedl1991} or understanding the discrepancy of multi-phonon UCN production in pressurized helium with calculations \cite{Schmidt-Wellenburg2015}, could provide valuable insights into this problem.

\section{Acknowledgements}

The authors would especially like to thank T. Soldner, the PF1B instrument responsible, for his time invested in the experiment and his contributions to this paper. We are also thankful for the help of: the entire cryogenic Sample Environment group at the ILL, especially E. Lelievre-Berna, O. Losserand, and X. Tonon; the Nuclear and Particle Physics group technicians M. Guigou, T. Brenner, and D. Berruyer, and well as other members of the group; C. Ismerie, summer internship student that helped during the development of the refrigerator; K. Andersen for sharing his inelastic neutron scattering data; and P. Huffman, R. Golub, and A. Young at North Carolina State University for their insightful comments on this paper.  KL's time for the preparation of this paper was supported in part by the US Department of Energy under Grant No. DE-FG02-97ER41042.

\section{Appendix: Temperature variation of UCN production}

Calculations of the relative UCN production rate $P_{\rm rel}(T)$ in Fig.~\ref{fig:relativeUCNproductionRate} are performed, following Refs.~\cite{Korobkina2002} and \cite{Schmidt-Wellenburg2009b}, using the expression
\begin{equation}
\label{eq:productionTemperatureCalcSLambda}
P(T) = N\sigma U_{\rm c} \frac{k_{\rm c}}{3\pi}\int^\infty_0 \frac{{\rm d}\phi}{{\rm d}\lambda_{\rm n}}\,s(\lambda_{\rm n})\,\lambda_{\rm n} \, {\rm d}\lambda_{\rm n} \;,
\end{equation}
where $N$ is the helium number density, $\sigma$ the bound neutron scattering length of $^4$He, $U_{\rm c}$ and $k_{\rm c}$ are the critical energy and momentum of UCN storable in the converter volume, $\lambda_{\rm n}$ is the incoming CN wavelength, and ${\rm d}\phi/{\rm d}\lambda_{\rm n}$ is the incident CN particle flux spectrum. The term $s(\lambda_{\rm n})$ is defined in terms of the dynamic structure factor of $^4$He $S(q,\omega)$, where $\hbar q$ and $\hbar \omega$ are the momentum and the energy transfers, via
\begin{equation}
s(\lambda_{\rm n}) \equiv \hbar \int S(q,\omega)\, \delta(\hbar\omega-E_{\rm n})\,d\omega \; ,
\label{eq:definitionOfslambda}
\end{equation}
where $E_{\rm n} = h^2/(2m_n\lambda_{\rm n}^2) = (\hbar k_{\rm n})^2/(2m_{\rm n})$, with $\hbar k_{\rm n}$ and $m_{\rm n}$ being the neutron momentum and mass.

\begin{figure}[tbp]
\begin{center}
\includegraphics[width=\columnwidth]{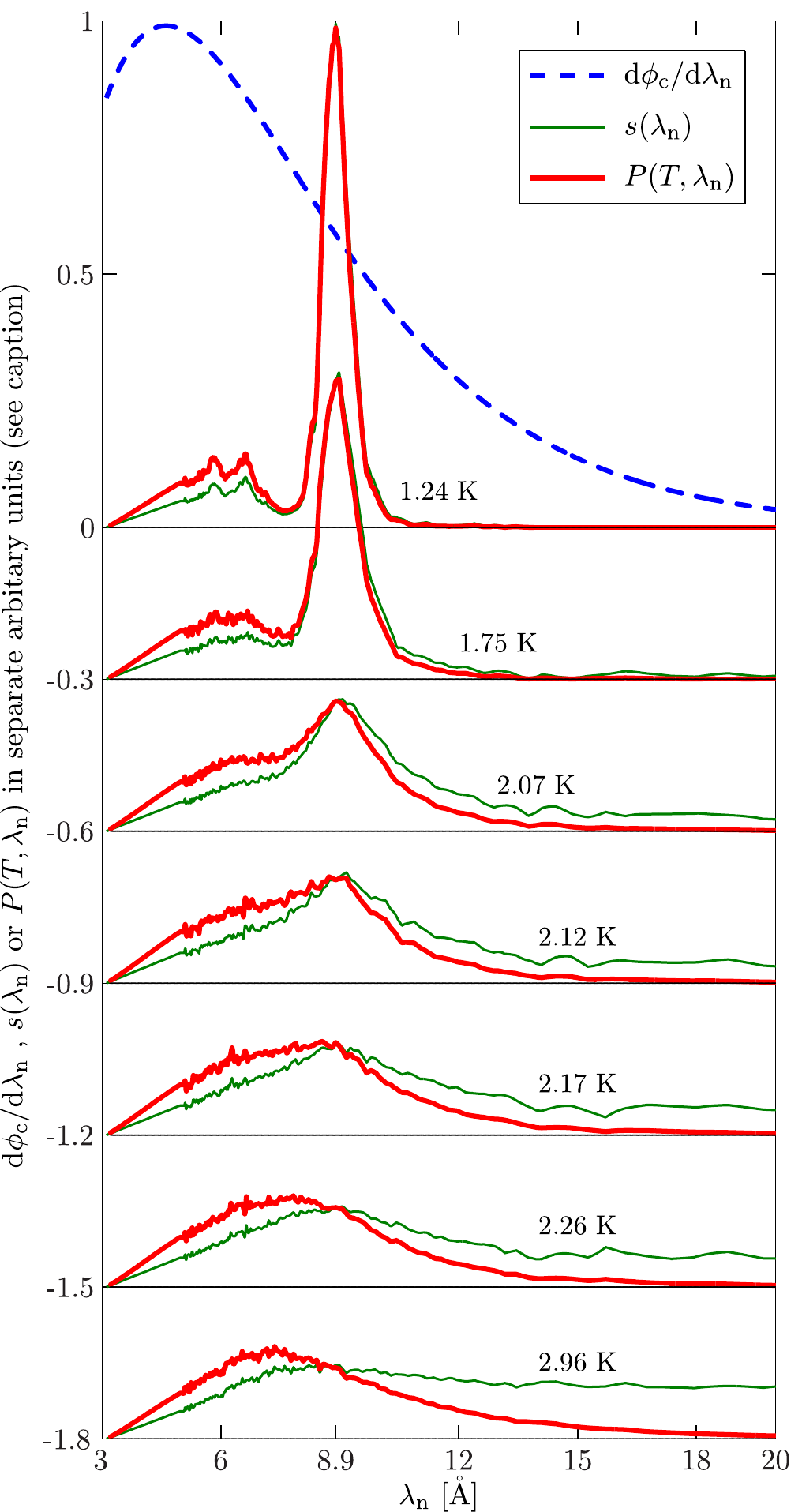}
\caption{(Color online) The PF1B CN capture flux ${\rm d} \phi_{\rm c} / {\rm d} \lambda_{\rm n}$, $s(\lambda_{\rm n})$ (Eq.~\ref{eq:definitionOfslambda}), and the differential UCN production rate $P(T,\lambda_{\rm n})$ used for the calculations of $P_{\rm rel}(T)$, at selected temperatures. Each separate quantity plotted is in the same arbitrary units for all temperatures, but the plots for incrementally increasing temperatures are offset by $-0.3$ (arbitrary units) from the previous. Details of the calculations are described in the text.}
\label{fig:UCNproductionWavelengthTemperature}
\end{center}
\end{figure}

The raw data \cite{Anderson2014} of $S(q,\omega)$ obtained from inelastic neutron scattering described in Ref.~\cite{Andersen1994a} is used in these calculations. These measurements were performed by scattering incident CN with a fixed energy of 3.8\un{meV} off high-purity liquid $^4$He in a 18\un{cm^3} container with the energy transfer determined using a time-of-flight neutron spectrometer. The measurements cover temperatures between 1.24\un{K} and 4.95\un{K} and $k$ between 0.3 and 2.1~$\text{\AA}^{-1}$. Plots of $s(\lambda_{\rm n})$ extracted from this data at selected temperatures are shown in Fig.~\ref{fig:UCNproductionWavelengthTemperature}.

The instrument energy resolution of these measurements is $\sim 100\un{\mu eV}$ (FWHM), which at the single-phonon peak, $\hbar\omega \approx 1\un{meV}$, corresponds to a $q$-resolution of $\sim 0.03\un{\text{\AA}^{-1}}$. This dominates the width of the single-phonon peak because, for example, the width of the single-phonon peak has been measured using neutron triple-axis and spin-echo spectrometers to be $\sim 7\un{\mu eV}$ \cite{Mezei1983}. When using a broad cold neutron spectrum, as is done here, the effect of the energy resolution is small when calculating $P(T)$.

Linearly extrapolating $s(\lambda_{\rm n})\rightarrow 0$ as $\lambda_{\rm n} \rightarrow 3\text{\AA}$ is done, and shown in Fig.~\ref{fig:UCNproductionWavelengthTemperature}. This is motivated by the measurements of Refs.~\cite{Fak1991} and \cite{Baker2003}. More sophisticated extrapolation schemes are available \cite{Schmidt-Wellenburg2009b}, but at the desired calculation precision, they are unnecessary.

The empirical approximation of the capture flux spectrum ${\rm d} \phi_{\rm c} / {\rm d} \lambda_{\rm n} = (\lambda_{\rm n}/1.8\text{\AA})\, {\rm d}\phi/{\rm d}\lambda_{\rm n}$ from Ref.~\cite{Abele2006} of the PF1B CN beam (also shown in Fig.~\ref{fig:UCNproductionWavelengthTemperature}) is used in these calculations. To compute the integral of Eq.~\ref{eq:productionTemperatureCalcSLambda} numerically, a constant spacing of ${\rm d}\lambda_{\rm n} = 0.01\un{\text{\AA}}$ was used. The $S(q,\omega)$ data points were 2D linear interpolated to calculate $s(\lambda_{\rm n})$.

Using the above, as well as the helium density \cite{VanSciver2012}, the differential UCN production rate $P(T, \lambda_{\rm n})$ is calculated and also shown in Fig.~\ref{fig:UCNproductionWavelengthTemperature}. The familiar structure: a clear separation of single-phonon and multi-phonon UCN production, can be seen at the lower temperatures. As the temperature increases, the single-phonon peak becomes broader and less prominent, and disappears beyond $T_\lambda$.

The $P_{\rm rel}(T)$ points in Fig.~\ref{fig:relativeUCNproductionRate} come from integrating $P(T, \lambda_{\rm n})$ over $\lambda_{\rm n}$ at different temperatures below 3.5\un{K}. The statistical errors of the $S(q,\omega)$ data \cite{Andersen1994a,Anderson2014} are combined to calculate the statistical error in $P_{\rm rel}(T)$. Since the spacing used to evaluate Eq.~\ref{eq:productionTemperatureCalcSLambda} and \ref{eq:definitionOfslambda} is smaller than the spacing of the $S(q,\omega)$ data, if a point is used multiple times, the errors are not combined independently (i.e. not added in quadrature). The statistical error in $s(\lambda_{\rm n})$ is $\sim 2\,\text{--}\,10\%$ in the regions that contribute significantly to UCN production. And the total statistical error in $P_{\rm rel}(T)$ is between 0.3$\,$--$\,$0.4\%.

\bibliographystyle{apsrev4-1}
\bibliography{SUN2PF1bpaper.bib}

\end{document}